\documentclass[acmsmall]{acmart}

\AtBeginDocument{%
  \providecommand\BibTeX{{%
    \normalfont B\kern-0.5em{\scshape i\kern-0.25em b}\kern-0.8em\TeX}}}

\setcopyright{acmcopyright}
\copyrightyear{2023}
\acmYear{2023}

\acmJournal{CSUR}
\acmVolume{37}
\acmNumber{4}
\acmArticle{111}
\acmMonth{12}
\usepackage{graphics}
\usepackage{enumitem}
\usepackage[english]{babel} 
\usepackage{blindtext}
\usepackage{pifont}
\usepackage{comment}
\newcommand{\cmark}{\ding{51}}%
\newcommand{\dquotes}[1]{``#1''}
\newcommand{\squotes}[1]{`#1'}
\usepackage{multirow}
\usepackage{tikz}
\usepackage{pgfplots}
\usepackage{array}
\usepackage{makecell}
\usepackage{xspace}
\usepackage{rotating}

\newcommand{\fars}{FARS\xspace}

\newcolumntype{L}[1]{>{\raggedright\let\newline\\\arraybackslash\hspace{0pt}}m{#1}}
\newcolumntype{C}[1]{>{\centering\let\newline\\\arraybackslash\hspace{0pt}}m{#1}}
\newcolumntype{R}[1]{>{\raggedleft\let\newline\\\arraybackslash\hspace{0pt}}m{#1}}
\usepackage{colortbl}%

\newcommand{\yashar}[1]{\textcolor{magenta}{{\bf [Yashar: }{\em #1}{\bf ]}}}
\newcommand{\alejandro}[1]{\textcolor{violet}{{\bf [Alejandro: }{\em #1}{\bf ]}}}

\definecolor{cos_colr}{rgb}{0.2, 0.7, 0.6}
\newcommand{\giovanni}[1]{\textcolor{cos_colr}{{\bf [Giovanni: }{\em #1}{\bf ]}}}

\newcommand{\argmax}{\operatornamewithlimits{argmax}}
\definecolor{blue(pigment)}{rgb}{0.2, 0.2, 0.6}
\def\updated#1{{\color{black}#1}}

\begin{document}

\title{A Review of Modern Fashion Recommender Systems}


\author{Yashar Deldjoo}
\authornote{Corresponding author, Email: deldjooy@acm.org}
\email{yashar.deldjoo@poliba.it}
\orcid{0000-0002-6767-358X}
\affiliation{%
  \institution{Polytechnic University of Bari}
  \country{Italy}
}

\author{Fatemeh Nazary}
\email{fatemeh.nazary@poliba.it}
\affiliation{%
  \institution{Polytechnic University of Bari}
  \country{Italy}
}

\author{Arnau Ramisa}
\authornote{Work done outside Amazon employment}
\email{aramisay@amazon.com}
\affiliation{%
  \institution{Amazon}
  \country{USA}
}

\author{Julian McAuley}
\email{jmcauley@eng.ucsd.edu}
\affiliation{%
  \institution{UC San Diego}
  \country{USA}
}

\author{Giovanni Pellegrini}
\email{g.pellegrini4@studenti.poliba.it}
\affiliation{%
  \institution{Polytechnic University of Bari}
  \city{Bari}
  \country{Italy}
}

\author{Alejandro Bellogin}
\email{alejandro.bellogin@uam.es}
\affiliation{%
  \institution{Autonomous University of Madrid}
  \city{Madrid}
  \country{Spain}
}

\author{Tommaso Di Noia}
\email{tommaso.dinoia@poliba.it}
\affiliation{%
  \institution{Polytechnic University of Bari}
  \country{Italy}
}

\renewcommand{\shortauthors}{Deldjoo et al.}

\begin{abstract}

The textile and apparel industries have grown tremendously \updated{over the last few years.} 
Customers no longer have to visit many stores, stand in long queues, or try on garments in dressing rooms as millions of products are now available in online catalogs. However, given the plethora of options available, an effective recommendation system is necessary to properly sort, order, and communicate relevant product material or information to users. Effective fashion RS can have a noticeable impact on billions of customers' shopping experiences and increase sales and revenues on the provider side.

The goal of this survey is to provide a review of recommender systems that operate in the specific vertical domain of garment and fashion products. We have identified the most pressing challenges in fashion RS research and created a taxonomy that categorizes the literature according to the objective they are trying to accomplish (e.g., item or outfit recommendation, size recommendation, explainability, among others) and type of side-information (users, items, context). We have also identified the most important evaluation goals and perspectives (outfit generation, outfit recommendation, pairing recommendation, and fill-in-the-blank outfit compatibility prediction) and the most commonly used datasets and evaluation metrics.

\end{abstract}




\maketitle

\section{Introduction}


Recommender systems 
help users navigate large collections of products to find items relevant to their interests
leveraging 
large
amounts of product information and user signals like product views followed or ignored items, purchases, or web-page visits to determine how, when, and what to 
recommend
to their customers. 
Recommender systems have grown to be an essential part of all large Internet retailers, driving up to 35\% of Amazon sales~\cite{mackenzie2013retailers} or over 80\% of the content watched on Netflix~\cite{chhabra17}. %

In this work, we are interested in recommender systems that operate in one particular vertical market: garments and fashion products. This setting introduces a particular set of challenges and sub-problems that are relevant for developing effective recommender systems.

Due to market dynamics and customer preferences, there is a large vocabulary of distinct fashion products,
as well as high turnover. This leads to sparse purchase data, which challenges the usage of traditional recommender systems~\cite{DBLP:journals/computer/KorenBV09}. Furthermore, precise and detailed product information is often not available, making it difficult to establish similarities between them.

To deal with the aforementioned problems, and given the visual and aesthetic nature of fashion products, there is a growing body of computer vision research addressing tasks like localizing fashion items~\cite{hara2016fashion, yamaguchi2012parsing}, determining their category and attributes~\cite{wang2018attentive, ferreira2019pose, chen2012describing}, or establishing the degree of similarity to other products~\cite{hadi2015buy, liu2012street, liu2016deepfashion, ak2018learning}, to name only a few.
Although 
works in the computer vision literature often don't consider
personalization (or recommendation), their predictions and embeddings can be leveraged by recommender systems and combined with past user preferences, thus mitigating sparsity and cold start problems.
Another relevant fashion problem that has attracted the attention of computer vision research is that of combining garments into complete outfits~\cite{DBLP:conf/mm/HanWJD17, lin2020fashion, bettaney2019fashion, chen2019pog}. Several works have studied how to learn the compatibility between fashion items using both professional photos of models wearing \updated{designer-created outfits} and social media pictures from `influencers' and normal people~\cite{vittayakorn2015runway, simo2015neuroaesthetics, hsiao2018creating}. In addition to allowing recommendations tailored to match the existing shopping basket or wardrobe of customers, these datasets help uncover other insights useful for recommender systems, such as the structure of fashion styles~\cite{hsiao2017learning}, social group preferences~\cite{kiapour2014hipster}, or the evolution of trends across time and location~\cite{StreetStyle2017,DBLP:conf/iccv/MallMHSB19}.


In addition to product-to-product relationships, fashion recommender systems also face particular product-to-user uncertainties, like \emph{fit}, that can hurt the quality of recommendations if not taken into account.
Fit prediction 
is a pain point for
online fashion shopping according to customers, and the primary reason for product returns faced by online retailers%
~\cite{cilley2016apparel}. The many sizing systems in use throughout the world, as well as their interpretation by different clothing manufacturers, make it very difficult to predict whether a particular product will properly fit a customer. Therefore, research on how to estimate a personalized garment fit has leveraged sources of information like co-purchase data~\cite{sheikh2019deep}, customer-reported measurements~\cite{parr2017impact}, or advanced imaging devices such as 3D scanners~\cite{ashdown2004using, apeagyei2010application, daanen20183d}. Needless to say, this information can be very valuable for recommendation.

Several online retailers, like Stitch-Fix and Amazon Prime Wardrobe, recreate the \emph{try-out} experience of brick-and-mortar stores by shipping a highly tailored assortment of garments to a customer, who can proceed to try them on in their home, and return those that they do not like. Recommender systems can make this approach more sustainable by maximizing the number of items that customers keep~\cite{zielnicki2019simulacra}, but the cold start problem and the commitment required from customers to stay in the membership plan make this approach hard to scale. Consequently, alternatives that would allow a customer to visualize the appearance and fit of clothes in an augmented reality or virtual environment are being researched~\cite{pachoulakis2012augmented}.

The reverse approach, namely designing or automatically generating a computer model of a garment with the right appearance and fit for a customer and subsequently manufacturing it, is another route being explored by, for example, Amazon with the \emph{Made for You Custom T-shirt}~\cite{madeforyou}, with the aim of bringing the made-to-measure fashion to the masses.

\subsection{Major Challenges}

In this section, we will describe the major challenges faced by recommender systems in the fashion domain.


\begin{description}
\item[Fashion item representation:] Traditional recommender systems such as Collaborative Filtering or Content-Based Filtering have difficulties in the fashion domain due to the sparsity of \updated{purchase data or the insufficient detail} about the visual appearance of the product in category names~\cite{DBLP:conf/waim/ShaWZFZY16}. Instead, more recent literature has leveraged models that capture a rich representation of fashion items through product images~\cite{DBLP:journals/corr/JagadeeshPBDS14,he2016vbpr}, text descriptions or customer reviews~\cite{DBLP:conf/sigir/ChenCXZ0QZ19, DBLP:conf/aaai/ZhaoHBW17}, or videos~\cite{yang2011real} which are often learned through surrogate tasks like classification or 
product retrieval. However, learning product representations from such input data requires large datasets to generalize well across different image (or text) styles, attribute variations, etc. Furthermore, constructing a representation that
learns which
product features customers take most into account when evaluating fashion products
is still an open research problem.

\item[Fashion item compatibility:] Training a model that is able to predict if two fashion items `go together' or directly combine several products into an outfit, is a challenging task. Different item compatibility signals studied in recent literature include co-purchase data~\cite{DBLP:conf/iccv/VeitKBMBB15,DBLP:conf/sigir/McAuleyTSH15}, outfits composed by professional fashion designers~\cite{DBLP:conf/mm/HanWJD17}, or combinations found by analyzing what people wear in social media pictures~\cite{DBLP:journals/mta/SunCWP18,DBLP:conf/recsys/Jaradat17}. From this compatibility information, associated image and text data \updated{are} then used to learn to generalize to stylistically similar products. Some works explicitly model the latent style types~\cite{DBLP:/corr/BracherHV16}. An additional under-explored difficulty for compatibility prediction is the dependency on trends, seasonality, location, or social group. 
Current approaches usually leverage image and text information.

\item [Personalization and fit:]  

The best fashion product to recommend depends on factors such as the location where the outfit will be used~\cite{chang2017fashion, al2020paris, wang2018analysis, kataoka2019ten}, the season or occasion~\cite{7907314, DBLP:conf/iccv/MallMHSB19, liu2017weather}, or the cultural and social background of the customer~\cite{kiapour2014hipster, DBLP:journals/mta/SunCWP18, zhong2019fashion}. 
A challenging task in fashion recommendation systems is how to discover and integrate these disparate factors~\cite{simo2015neuroaesthetics, 8455311}. Current research often tackles these tasks by utilizing large-scale social media data.

As discussed earlier, a personalization dimension very particular to the fashion domain is that of fit. In addition to predicting what size of a product will be more comfortable to wear, body shape can influence stylistic choices~\cite{sattar2019fashion, DBLP:conf/cvpr/HsiaoG20, DBLP:conf/mm/HidayatiHCHFC18}.

\item [Interepretability and Explanation:] 

Most of the existing fashion recommender systems in the literature focus
on improving  predictive performance, treating the model as a black box.
However, deploying accountable and interpretable systems able to explain their recommendations can foster user loyalty in the long term and improve the shopping experience. Current models generally offer explanations through highlighted image regions and attributes or keywords~\cite{wu2020visual, DBLP:conf/ijcai/HouWCLZL19, packer2018visually, yang2019interpretable, han2019prototype, liao2018interpretable}.

\item [Discovering Trends:] 
Being able to forecast consumer preferences is valuable for fashion designers and retailers in order to optimize \updated{product-to-market fit}, logistics, and advertising.
Many factors are confounded in what features are considered \squotes{fashionable} or \squotes{trendy}, like seasonality~\cite{DBLP:conf/iccv/MallMHSB19}, geographical influence~\cite{al2020paris}, historical events~\cite{hsiao2021culture}, or style dynamics~\cite{al2017fashion, lo2019dressing, furukawa2019visualisation}. Again, social media is a useful resource leveraged by researchers~\cite{ding2021leveraging, huang2021street}.

\end{description}
The challenges above, as well as many other issues have been discussed in the studied research works, and they are reviewed in the following sections. For instance, visual modeling of fashion items (cf. Section~\ref{sec:UI_side_info}), modeling of fashion outfits (cf. Section~\ref{subsec:outfit_rec}), geo-temporal localization (cf. Section~\ref{subsec:context}), attribute prediction for fashion and leveraging multi-modal data  (cf. Section~\ref{sec:UI_side_info}), explainability in fashion recommendation (cf. Section~\ref{subsec:expl}) and learning style (cf. Section~\ref{sec:UI_side_info}).

\subsection{Search strategy for relevant papers}\label{subsec:howCollect}

We primarily relied on papers indexed in DBLP,\footnote{\url{https://dblp.uni-trier.de/}} a prominent computer science bibliographic database, to identify publications that comprise the state-of-the-art in fashion item and outfit recommendation. Our search approach was divided into two stages: finding relevant publication collections; and (ii) post-processing and filtering the final list. The total number of articles analyzed exceeds 50, with the majority of recognized research papers published between 2014 and 
\updated{2023}, 
an indication of the topic's originality and freshness. While we do not claim that this study is exhaustive in terms of the collected papers, we believe it gives a comprehensive overview of current achievements, trends, and what we deem to be the most significant challenges and tasks in fashion recommender systems. Additionally, it gives valuable information to both industry practitioners and academics.\footnote{This link contains a comprehensive list of papers, which will be periodically updated: \url{https://github.com/yasdel/FashionXrecsys}}

\subsection{Related Surveys}
To put this survey in context, we identified and present 
related review and survey articles to explain in which ways our article differs from and extends earlier work. 
In a recent work, \citet{deldjoo2020recommender} presented a survey of RS leveraging multimedia content, i.e., visual, audio, and/or textual features. The domains studied in this survey include various ones such as media streaming for audio and video recommendation, e-commerce for recommending different products including fashion items, news, and information recommendation, social media, and so forth. While fashion RS were also discussed,
the authors only included a small portion of the topics and papers in this domain. Here, we discuss and present a comprehensive survey of significant tasks, challenges, and types of content used in the fashion RS field. 

We have also identified surveys~\cite{cheng2021fashion,zoghbi2016fashion} where the authors present a literature review of techniques at the intersection of fashion and computer vision (CV) and/or natural language processing (NLP). While we find these works relevant to this article, they remain largely different from the review presented here as those systems are not focused on RS but on other aspects of the fashion domain, such as text generation from images or pose estimation. 

Perhaps the most relevant work to our current survey is a recent book chapter by~\citet{fashion_rshb_2021} on fashion RS. 
This chapter focuses on discussing the state of the art of fashion recommendation systems; in particular, the authors affirm that deep learning represented a turning point with respect to the canonical approaches and therefore the authors examined four different tasks that use this new approach. Additionally, they provided examples and possible problems and their evaluation.
In particular, the authors focused their review on tasks related to social media and the size recommendation problem (see Section~\ref{sss:size_rec}, where we introduce this task in detail). 
In our survey, in addition to analyzing the state of the art of the most commonly used algorithms in a wide range of tasks, we went in-depth to understand which are the main features used by the more modern fashion recommender systems.
In fact, an extensive discussion is held on how both the user and the items, with their characteristics, can be a source for the definition of models with accurate recommendations.
Furthermore, since the fashion domain 
focuses on
visual aspects, a concise reference is made regarding the utilization of computer vision techniques in the realm of fashion recommendation systems.


\noindent \updated{\textbf{The outline of the survey.} The structure of the paper is organized according to the following:}
\begin{itemize}
    \item \updated{Section~\ref{sec:cat_fars} is dedicated to the \dquotes{\textit{categorization of fashion recommender systems}}, which we organize according to the \textbf{task} (cf. Section~\ref{subsec:fars_task}) such as fashion item recommendation, pair, and outfit recommendation, sizer recommendation, among others, based on the \textbf{input} (cf. Section~\ref{subsec:fars_data_input}), e.g., by relying on collaborative information, textual and visual features, etc.}
    
    \item \updated{Section~\ref{subsec:algs} is dedicated to the \dquotes{\textit{algorithms for fashion recommender systems}}, which we perform based on \textbf{visually-aware CF} (cf. Section~\ref{subsub:VRS}), \textbf{generative models} (cf. Section~\ref{subsec:GAN_FA_rec}), other models  (cf. Section~\ref{subsec:others}), where the preceding two sections also touch on topics related to  \textbf{computer vision} and fashion recommendation systems in a brief manner.}
    

    \item \updated{Section~\ref{sec:eval} is dedicated to 
    \dquotes{\textit{evaluation and datasets}}, which we discuss based on \textbf{evaluation goal} (cf. Section~\ref{subsec:eval_goal}), \textbf{evaluation perspective} (cf. Section~\ref{subsec:eval_pers}) discussing 
    themes such as whether an evaluation is intended to evaluate recommendation, explanation, and image quality, and finally \textbf{available datasets} (cf. Section~\ref{subsec:avail_data}).}

    \item \updated{Finally, Section~\ref{sec:con} discusses \dquotes{\textit{conclusion and future challenges}}.}


\end{itemize}

\updated{In Table~\ref{tbl:abbr}, we mention the list of abbreviations used throughout this paper.}

\begin{table}[t]
\centering
\caption{List of abbreviations used throughout this paper.} 
\label{tbl:abbr}
\scalebox{0.80}{

 \begin{tabular}{l|l}
\toprule

\multicolumn{1}{c}{\textbf{Abbreviation}} & \multicolumn{1}{c}{\textbf{Term}} \\

\bottomrule
AI & Artificial Intelligence \\ \hline
AML & Adversarial Machine Learning \\ \hline
AMR & Adversarial Multimedia Recommendation \\ \hline
AUC & Area Under the ROC Curve \\ \hline
BPR & Bayesian Personalized Ranking \\ \hline
C\&W &Carlini and Wagner \\\hline
CA & Context-Aware RS \\ \hline
CBF-RS & Content-Based Filtering RS\\ \hline
CF-RS & Collaborative Filtering RS \\ \hline
CNN & Convolutional Neural Network \\ \hline
CS & Cold start \\ \hline
CD-RS & Cross-Domain RS\\ \hline
CV & Computer Vision \\ \hline
DL & Deep Learning \\ \hline
DNN & Deep Neural Network \\ \hline
ERM & Empirical Risk Minimization \\ \hline
FCN & Fully Convolutional Network \\ \hline
FGSM & Fast Gradient Sign Method  \\ \hline
FTF & Functional Tensor Factorization \\ \hline
GAN & Generative Adversarial Network \\ \hline
\bottomrule
\end{tabular}

\quad

 \begin{tabular}{l|l}
\toprule

\multicolumn{1}{c}{\textbf{Abbreviation}} & \multicolumn{1}{c}{\textbf{Term}} \\

\bottomrule

G-RS & Graph-based RS \\ \hline
GRU & Gated Recurrent Unit \\ \hline
IR & Information Retrieval \\ \hline
KG & Knowledge graph \\ \hline
K-NN & k-Nearest Neighbors \\ \hline
LFM & Latent Factor Model \\ \hline
LSTM & Long Short-Term Memory \\ \hline
MF & Matrix Factorization \\ \hline
ML & Machine Learning \\ \hline
NLP & Natural Language Processing \\ \hline
RNN & Recurrent Neural Network\\ \hline
ROC & Receiver Operating Characteristic \\ \hline
RS & Recommender Systems \\ \hline
SGD & Stochastic Gradient Descent \\ \hline
SM & Social Media \\ \hline
SN & Social Network \\ \hline
SVM & Support-Vector Machines \\ \hline
TF-IDF & Term Frequency–Inverse Document Frequency \\ \hline
URM & User Rating Matrix \\ \hline
VRS & Visual recommender system \\ \hline

\bottomrule
\end{tabular}

}
\end{table}

\section{Categorization of Fashion Recommender Systems}
\label{sec:cat_fars}

We categorize fashion recommender systems according to their task  (cf. Section~\ref{subsec:fars_task}), and the input data they use to perform that task (cf. Section~\ref{subsec:fars_data_input}). We will discuss these two categorizations in more detail in the next sections.

\subsection{Categorization based on task}\label{subsec:fars_task}
By task, we refer to the internal goal the fashion RS aims to achieve. This affects, in particular, the expected output of the algorithms. We identified five main tasks in the academic literature that fashion RS aim to achieve:
\textit{(i)} Fashion item recommendation,
\textit{(ii)} Fashion pair and outfit recommendation, \textit{(iii)} Size recommendation, \textit{(iv)} Explanation for Fashion Recommendation and \textit{(v)} Other fashion prediction tasks. 
There are a few other tasks that so far have not gathered much attention, but are growing in popularity; they will be summarized later under the general subsection (v).

We discuss these \updated{five} categories by presenting a broad definition of each task formally while leaving a more detailed discussion of several prominent research works for Section~\ref{subsec:algs}


\subsubsection{Fashion item recommendation:} The fashion item recommendation task, similar to the classical recommendation problem, focuses on suggesting individual fashion items (clothing), that match users' preferences. 

\begin{definition}[Fashion item recommendation]
Let $\mathcal{U}$ and $\mathcal{I}$ denote a set of users and fashion items in a system, respectively. Each user $u \in \mathcal{U}$ is related to $\mathcal{I}_u^+$, the set of items she has consumed. Given a utility function $g : \mathcal{U} \times \mathcal{I} \rightarrow \mathbb{R}$ the \textbf{Item Recommendation Task} is defined as
\begin{equation}
\label{eq:item_util_gen}
    \forall u \in \mathcal{U}, \ i^{*}_u = \argmax_{i\in \mathcal{I} \setminus \mathcal{I}_u^+} g(u,i)
\end{equation}
where $i^{*}_u$ denotes the best matching item not consumed by the user $u$ before. 
The preference of user $u$ on item $i$ 
could be encoded as $s_{ui} \in \mathcal{S}$, a continuous-value score (e.g., 1-5 Likert scale), or implicit feedback in which we assume the user likes the item if she has interacted with
(i.e., reviewed, purchased, clicked)  the item (i.e., $s_{ui} = 1$). 
$\mathcal{I}_u^+$ represents the set of $(u, i)$ pairs for which $s_{ui}$ is known. The task of personalized Top-$K$ fashion item recommendation problem is formally defined as identifying, for user $u$, a set of ranked list of items $X_u=\{i_1, i_2, ..., i_K\}$ that match user preference. 

\end{definition} \qed


A simple yet effective pure CF model that serves as a foundation for many model-based VRS is BPR-MF~\cite{DBLP:conf/uai/RendleFGS09}.
Given a user $u$, and an item $i$, the core predictor in this model is given according to:
        \begin{equation}
        \label{eq:MF}
        \hat{s}_{u, i} =  p_u^Tq_i
    \end{equation}
where $p_u, q_i \in \mathbb{R}^{F}$ are the embedding vectors for user $u$ and item $i$, respectively, and $F$ is the size of the embedding vector. This predictor is known as Matrix Factorization (MF),
the parameters of the model can be learned using Bayesian Personalized Ranking (BPR), a pairwise ranking optimization framework. We will show later in Section~\ref{subsub:VRS} how different authors have used this model as a starting point in the fashion domain so that it is extended to consider other types of inputs and learning scenarios.


\subsubsection{Fashion pair and outfit recommendation:}\label{subsec:outfit_rec}
Fashion outfits are sets of $N$ items worn together, e.g., for an outdoor wedding, graduation party, baby shower, and so forth. The simplest form of a fashion outfit is when $N=2$,
i.e., two different 
items
that look good when
paired together, such as an orange shirt and a blue pair of jeans. However, in general, outfits in online fashion stores can be composed of items in different categories (e.g., show, bottom, top, hat, bag) that share some stylistic relationship.

\begin{definition}[Fashion outfit composition] 
Let $O = \{i_{1}, i_{2}, ..., i_{N}\}\in \mathcal{O}$ denote a fashion outfit composed of a set of compatible fashion items, in which $i_{n} \in \mathcal{I}$ is the $n$-th fashion item in the outfit $O$, and $\mathcal{O}$ is the set of all possible outfits, $N$ is the length of the outfit, whose value is not fixed and can change with each outfit. Fashion Outfit Composition is formulated as follows: \dquotes{given a scoring function $s(O)$ 
 that indicates how well the outfit $O \in \mathcal{O}$ is composed, find an outfit $O$ that maximizes this utility}: 
\begin{equation}
\label{eq:fashion_outfit_comp1}   
O^{*} = \argmax_{O_j \in \mathcal{O}} s(O_j)
\end{equation}
where $s$ is the outfit utility function, a scoring function that takes into account different types of relationships between fashion items to generate an outfit compatibility score. It is worth noting that, given this general task definition, the \textit{evaluation} of fashion outfit composition is typically performed via fill-in-the-blank (FITB) or outfit compatibility score prediction~\cite{DBLP:conf/icip/PolaniaG19} described in Section \ref{sec:eval}.

\end{definition}
\qed


Moreover, it is possible to encode several objectives relevant to the fashion domain in the definition of the outfit composition scoring function by incorporating domain knowledge. 
For instance,~\citet{DBLP:conf/sigir/McAuleyTSH15} defined compatibility according to \textit{complementary} relationships (e.g., how much a white shirt complements blue pants), and \textit{similarity} (how much one item in the outfit is visually similar to another item).~\citet{hsiao2018creating} model the outfit scoring function $s(O_j)$ as the superposition of two objectives:
\begin{equation}
\label{eq:fashion_outfit_comp2}   
O^{*}_j = \argmax_{O_j \in \mathcal{O}} c(O_j) + v(O_j)
\end{equation}
\noindent where $c(.)$ and $v(.)$ denote the \textit{compatibility} score (how much pairs complement each other) and \textit{versatility} score (defined as coverage of all styles), respectively. Parameters of the outfit scoring functions may be personalized to each style/user, and learned from user interaction data. As an example, in~\cite{lin2020fashion} the authors propose a system where compatibility is computed according to the image and category of every item in the outfit. 

Recommending a pair of fashion items or an outfit evidences several challenges, which have been addressed as follows:

\begin{itemize}
    \item \textbf{Personalizing to a target 
    customer.}
    Compatible fashion outfit  recommendations can be studied in a personalized fashion with respect to the target user $u$ to whom recommendations are computed:
\begin{equation}
\label{eq:fashion_outfit_comp1_person}   O_u^{*} = \argmax_{O_j \in \mathcal{O}} s(u, O_j).
\end{equation}
A common problem here is the availability of sufficient interaction data to learn personalized outfit models. Different approaches have been proposed to address this issue in the literature, notably by using fashion generation~\cite{chen2019pog}, graph-learning approaches~\cite{DBLP:conf/sigir/LiW0CXC20}, binary codes~\cite{DBLP:conf/cvpr/Lu0JCZ19}, and finally self-attention mechanism to model the higher-order interactions between fashion items~\cite{lu2021personalized}.


 
   \item \textbf{Modeling outfits as a sequence.} To take advantage of the representation of order-aware models such as LSTMs, some works~\cite{DBLP:conf/mm/HanWJD17}  model an outfit as an ordered sequence of items, where each item belongs to step $t$. Then sequence-aware methods model the visual compatibility relationships of outfits as a next item recommendation task:
   \begin{equation}
    \forall O_t = \{x_1, x_2, ...,x_{t}\}, \ x_{t+1} = \argmax_{x \in \mathcal{Y}} s(O_t,x)
    \end{equation}
    where $\mathcal{Y}$ contains the set of allowed items. For instance, a user may want to find the best \dquotes{shirt} that matches his/her current \dquotes{shoes} and \dquotes{pants}. In this example, all the \dquotes{shirts} in the database could be a candidate item set, thus filling the cycle: \dquotes{shoes $\rightarrow$ pants $\rightarrow$ shirts}, and this cycle can go on to add more items such as \dquotes{shoes $\rightarrow$ pants $\rightarrow$ shirts $\rightarrow$ hat $\rightarrow$ bag} (bottom-up outfit representation), depending on the size of outfit considered. 
    
    This problem might be generalized by allowing an input query $x_q$, either as a textual query (e.g., what outfit goes well with this mini skirt?) or a visual query (photo of a skirt). 
    Moreover, when $x_q$ is a visual query, the task is recognized as a \textbf{visual retrieval} task in the IR community or \textbf{contextual recommendation} in the RS community.
    
    \item \textbf{Content-based versus collaborative filtering.} While fashion item recommendation tasks have been 
    dominated by CF-based approaches, which may or may not include side visual or textual information, the problem of outfit recommendation has been predominantly approached via content-based approaches, e.g., via visual metric learning problems. This can be explained due to scarcity of outfit interaction data. Nevertheless, there have been few works that approach the outfit problem in a CBF-driven approach consider the work by \citet{DBLP:conf/mm/HuYD15}, in which the authors use multi-modal features of items and tensor factorization to model the interactions between users and fashion items.
\end{itemize}

Here, we review some notable research works that study fashion outfit composition tasks. Two crucial issues arise here~\cite{DBLP:conf/www/YinL0019}:

\begin{itemize}
    \item How can we learn domain knowledge about fashion compatibility relationships between fashion items in the presence of subtleties and subjectivity that can exist across customers?
    \item How can we incorporate the learned domain knowledge into the design space of fashion recommendation models?
\end{itemize}
A number of works have tried to model visual/style compatibility as a \textit{metric learning} problem, for instance, for clothes~\cite{DBLP:conf/icdm/HePM16,DBLP:conf/sigir/McAuleyTSH15,DBLP:conf/icip/PolaniaG19}, and for furniture~\cite{DBLP:conf/dagm/AggarwalVSY18,DBLP:journals/tog/Bala15,DBLP:journals/mta/PanDHC19}. \citet{DBLP:conf/www/YinL0019} propose a fashion recommendation system that, in the first phase, learns and incorporates visual compatibility relationships of items at the pixel level. In the second phase, a domain adaptation strategy is used to allow the use of the knowledge extracted from the source domain that can be consistent with a different domain, that is, how to make sure that the knowledge learned in the database used in their experiments can also be compatible with others.

\citet{DBLP:conf/icip/PolaniaG19} proposed a neural model composed of two sub-networks, first a Siamese sub-network which extracts visual features from a pair of images and a metric learning approach that maps the pair of features and auxiliary cues (color information) to a fashion compatibility score.
\citet{DBLP:journals/tmm/LiCZL17}  proposed learning a representation of an outfit by exploiting the information from the image, title, and category and a multi-modal fusion model. Prediction is made by computing a compatibility score for the outfit representation. The scoring component is learned in an end-to-end fashion. \citet{DBLP:conf/mm/HanWJD17} model the outfit generation process as a sequential process and employ a bidirectional LSTM to predict the next item from the current one. However, these works build on two key assumptions: \textit{(i)} a fixed number
of items in an outfit or \textit{(ii)} a fixed order of items in terms of their
categories, which may not exist in reality. To address this issue, other approaches have been proposed that do not build on these assumptions.  In order to address the issue of outfit recommendation,~\citet{DBLP:conf/www/LinMY20} employ a multiple instance learning (MIL) approach. This assumes that labels for a group of items are known, but not for each individual item. They propose a two-stage framework called OutfitNet for a fashion outfit selection. In the first stage, a Fashion Item Relevancy network (FIR) is used to learn the compatibility of fashion items and garment item relevance embeddings. In the second stage, an Outfit Preference network (OP) uses visual input to learn consumers' preferences. OutfitNet takes in various fashion products in an outfit, learns the compatibility of the fashion items, the users' preferences for each item, and the users' attention on different pieces in the outfit.

\updated{Recent works in the fashion domain propose scalable approaches for personalized outfit recommendations. \citet{chen2019pog} present an industrial-scale fashion recommender system, POG, that uses a multi-modal neural model and transformer architecture to better capture dependencies among items. \citet{sarkar2023outfittransformer} propose OutfitTransformer, a scalable framework that uses self-attention to learn relationships between items in an outfit modeled as an unordered set. The system is trained using a novel set-wise outfit ranking loss, which generates a target item embedding and specification based on an outfit.}

\subsubsection{Size Recommendation.}
\label{sss:size_rec} 
\updated{As the fashion industry continues to evolve, it has become increasingly important to consider more subjective factors such as fit, fabric softness, and material smell in fashion recommendation systems. According to \cite{jaradat2020second}, the ability to take measurements with mobile phones and apply algorithms will become crucial in this domain. However, the issue of size-fit remains a key concern for customers, leading to significant product returns and financial losses for companies. To address this challenge, size recommendation systems have been developed to help users select the best-fitting items without physically trying them on. Despite these efforts, determining the appropriate size and fit for customers still poses various challenges, as highlighted by \citet{fashion_rshb_2021}.}




\begin{description}
\item [Lack of consistency between brands.] There is a large number of approved sizing systems around the globe for various clothes, such as dresses, tops, skirts, pants and brands. Moreover, there are different size systems such as numeric (38-39-40), standard (S,M,L), fractions (41 1/3, 42.5), convention sizes (36-38, 40-42), country conventions (EU, FR, IT, UK), where inconsistencies and different ways of converting a local size system to another (as brands do not always comply with the same conversion logic) make the task challenging.

\item [Subjectivity.] The {exact size} is a very subjective feature; users who have purchased items with the same style and shape may make future purchases with different sizes; how an item fits on your body depends on or can be influenced by several factors, making an objective recommendation difficult. Moreover, customers may be driven by emotional aspects; even a piece of accurate size advice can come with a high emotional cost when the advised size differs from the customer's expectation.

\item [Data sparsity.] Users are able to buy only a small part of the items of an e-commerce website and on the other hand, articles have a limited stock, which can in turn hinder the task of recommender systems working with user-item fit feedback.
\item [Noise.] The underlying interaction data where fashion recommenders are trained can be very noisy since users may make purchases not only for themselves but also share information about clothing with their friends and family, hindering the task of size recommendation.

\end{description}
Given a fashion garment and the shopping history of a user, size recommendation methods predict whether a given size will be too large, small, or correct. Approaches for size recommendation can be classified according to which type of input data they rely on:
\begin{itemize}
    \item \textit{Physical body-related features.} The easiest way to make effective sizing recommendations is to use data from certain parts of the body~\cite{DBLP:conf/cvpr/HsiaoG20, DBLP:conf/mm/HidayatiHCHFC18} such as bust, waist, and hip. Sometimes it may be useful to involve the user directly, for example by providing questionnaires to obtain other information, such as age, sex, height, and weight, or in more recent systems to provide their own photos and through vision algorithms to extract 3D avatars that reflect the physical characteristics of the users~\cite{kim2016unity3d}. 
    \item \textit{User-item fit feedback.} To provide personalized size recommendations, the interaction between the user and the item is essential; in this sense, using purchase data~\cite{DBLP:conf/www/AbdullaSB19} or user returns and features of items~\cite{DBLP:conf/recsys/GuigouresHKSBS18, DBLP:conf/cvpr/HsiaoG20, DBLP:conf/mm/HidayatiHCHFC18} are used for this purpose, while allowing the user not to have to provide 
    data relating to their physical appearance. 
\end{itemize}
 

A few research works on size recommendation are reviewed here.
\citet{DBLP:conf/www/AbdullaSB19} propose a size recommender system that treats the recommendation problem as a classification problem. It learns a joint size-fit shared latent space to project users and fashion items into, using the skip-gram-based Word2Vec model. A gradient-boosting classifier is then used to predict the fit likelihood. The authors validate the usefulness of their system with both offline and online evaluation. 

\updated{The implementation of Bayesian models is described in 
\cite{DBLP:conf/recsys/GuigouresHKSBS18,DBLP:conf/kdd/NestlerKHWS21} to address size and fit-related returns, which involve keeping, returning, or exchanging an article due to sizing issues. These models utilize prior knowledge and expertise to enhance their accuracy and have been tested on a massive scale of data from millions of customers. In summary, these models exhibit significant potential in reducing returns and enhancing customer satisfaction in the fashion e-commerce industry.}

\updated{In the fashion industry, choosing clothing that fits well and flatters one's \textbf{body shape} is crucial. To address this issue, two approaches have been discussed below. The first, VIBE (short for VIsual Body-aware Embedding) proposed by~\citet{DBLP:conf/cvpr/HsiaoG20} is a framework that captures clothing's association with various body shapes. VIBE uses an embedding learned from images of fashion models of different shapes and sizes wearing various garments to recommend clothing that will suit the user's body shape. This approach allows for the creation of a 3D model of the human body that can be used for personalized recommendations. The second approach, proposed by \citet{DBLP:conf/mm/HidayatiHCHFC18}, is a clustering-based body shape assignment approach that compares the body measurements of celebrities to assign them to body shape categories such as hourglass, rectangle, round, and inverted triangle. The system generates networks of images based on the visual appearance of clothing styles and body shape information, which are then used to recommend clothing for each body shape. However, one issue with this approach is that the body shapes of celebrities may not correspond to those of typical users. Overall, both approaches offer promising methods for personalized clothing recommendations that take body shape into account.}

\begin{figure}[t]
    \centering
    \includegraphics[width = 0.8\linewidth]{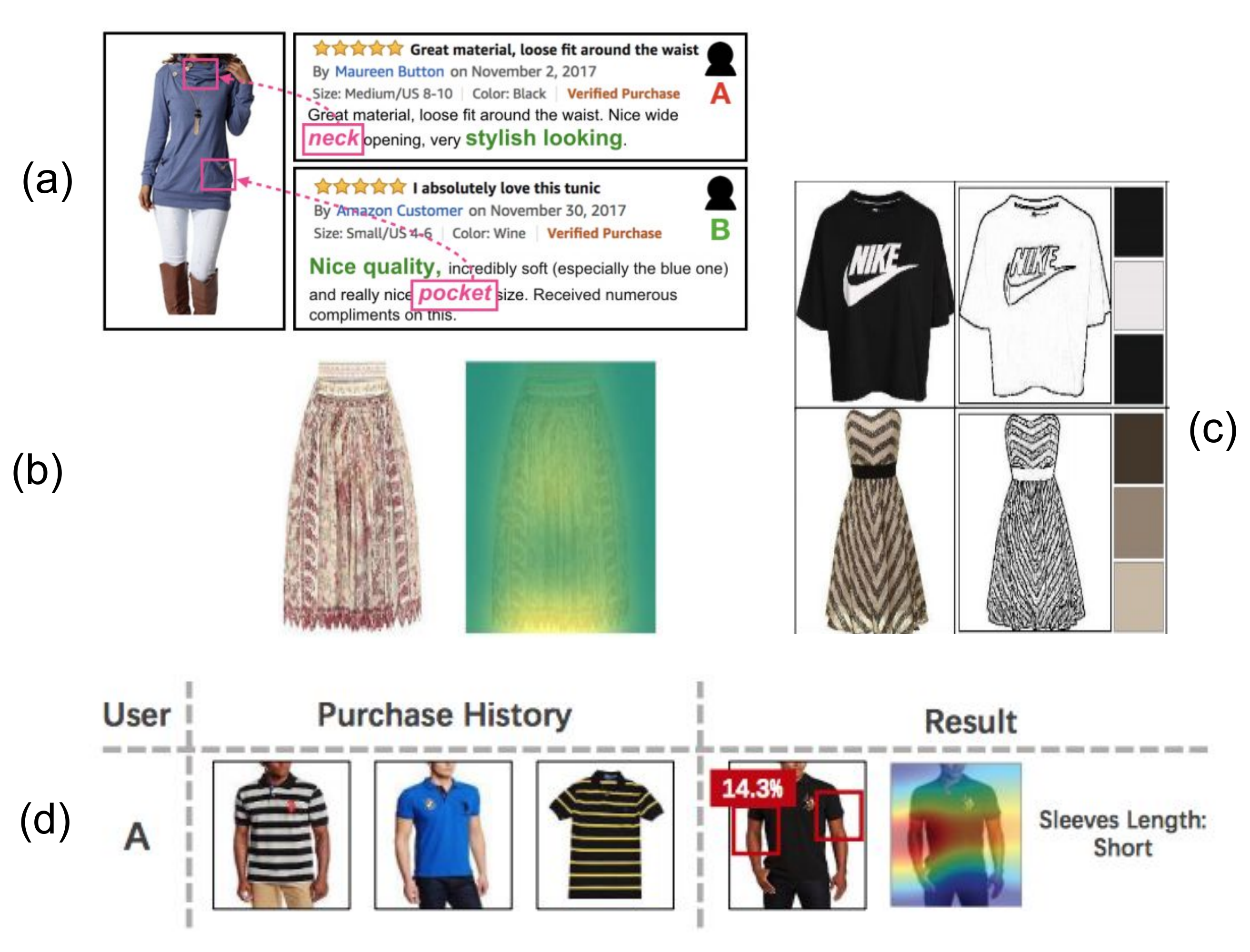}
    \caption{(a) Personalized visual explanation by \citet{DBLP:conf/sigir/ChenCXZ0QZ19} related to region of the fashion image based on user reviews; (b) Explainable outfit recommendation by \citet{DBLP:journals/tkde/LinRCRMR20} using a neural attentive model; (c) outfit recommendation  by \citet{DBLP:conf/wacv/TangsengO20}, and (d) fashion item recommendation proposed by \citet{DBLP:conf/ijcai/HouWCLZL19} comparing recommendation results with the target user's purchase history.}
    \label{fig:my_label}
\end{figure}




\subsubsection{Learning explanations for fashion recommendation.}\label{subsec:expl} An explanation is a piece of information that is shown to users that describes why a particular item is recommended. In recent years, there has been an increase in the number of studies on \textit{explainable recommendation} in various areas, especially the fashion domain, which can be categorized along the following dimensions:



\begin{itemize}
    \item \textbf{Explanation goal.} Explanations have a significant impact on how individuals react to recommendations. A RS's aims in creating explanations for a particular recommendation, however, might range from enhancing the transparency to facilitating a speedier decision to convincing the user~\cite{tintarev2015explaining,balog2020measuring}.
    \item \textbf{Information source for explanation.} Recommendation explanations can be derived from many information sources and be presented in a variety of display styles. Examples of such styles in the fashion domain include textual explanations~\cite{cheng2019mmalfm,DBLP:conf/sigir/ChenCXZ0QZ19}, or visual explanations to specific regions in the image~\cite{DBLP:conf/sigir/ChenCXZ0QZ19}, feature-level explanations based on fashion items' content features~\cite{DBLP:conf/wacv/TangsengO20,deldjoo2022leveraging}, and by comparison with the user purchase history~\cite{DBLP:conf/ijcai/HouWCLZL19}.
    \item \textbf{Explanatory \updated{approach/focus}:} \updated{Explainable recommendation research can focus on either developing interpretable~\textit{models} for increased transparency  or developing separate post-hoc/model-agnostic approaches to explain recommendation~\textit{results}. Thus, the contrast lies in whether the explainability is focused on the method itself or the outcome.} 
    
\end{itemize}

\updated{In other words, the explanation goal asks, \dquotes{What is the explanation's ultimate objective?} whereas the explanatory approach/focus seeks to comprehend \dquotes{What will the explanation focus on?}}. In Figure~\ref{fig:my_label}, different types of visual explanation used for fashion recommendation are illustrated.~\citet{DBLP:conf/sigir/ChenCXZ0QZ19} present a method that provides visual explanations to the user through certain regions of the items, where an LSTM combines visual information and user reviews. With the same goal~\citet{DBLP:conf/wacv/TangsengO20} propose a system that provides outfit recommendations and provides scores for each item or for each item feature in order to understand how much these influence the outfit composition. To explain whether an item/feature is compatible with the outfit and its influence on the score, three attributes have been extracted from the images: shape, texture, and colors. K-means clustering is applied for colors, and a CNN is used for shape and texture representation.

\subsubsection{Other fashion-related prediction tasks}
Besides fashion items and outfit recommendations, a few similar tasks are gaining attention in the fashion and retail industry. One of these areas includes \textit{image-based fashion generation}, where methods have been proposed to generate outfits so that the user could explore them and judge their compatibility~\cite{DBLP:conf/icdm/KangFWM17,yang2018recommendation,DBLP:conf/www/LinRCRMR19}. One of the key techniques used in these works is Generative Adversarial Networks (GANs), which will be described in detail in Section~\ref{subsec:GAN_FA_rec}.
Other relevant prediction tasks concern the fashion AI such as design~\cite{DBLP:conf/icdm/KangFWM17}, wardrobe creation~\cite{hsiao2018creating}, fashion trend forecasting~\cite{al2017fashion}, and societal/marketing biases such as socio-demographic
inequality structures through recommendation systems in fashion~\cite{DBLP:conf/hcse/BrandG20}, or user shopping behavior 
online and in brick-and-mortar stores~\cite{DBLP:conf/um/WolbitschHWH20}, or capturing users' seasonal and temporal preference changes~\cite{DBLP:conf/fsdm/XingHS20}. 





\begin{table}
\centering
\caption{Classification of different research works incorporating various side information kinds in their proposed approach.}
\label{tbl_feats}
\resizebox{\linewidth}{!}{%
\begin{tabular}{l|l|c|c|l|l|l|l|l|l|l|l|l|l|l|l|l|l|l|l|l|l|} 
\hline
\multicolumn{1}{c|}{\textbf{Authors}} & \textbf{Year} & \multicolumn{20}{c|}{\textbf{Data/computational features}} \\ 
\hline
\multicolumn{1}{l}{} &  & \multicolumn{2}{c|}{Side-info User} & \multicolumn{14}{c|}{Side information of items} & \multicolumn{4}{c|}{Other information} \\ 
\cline{3-22}
\multicolumn{1}{l}{} &  &  &  & \multicolumn{7}{c|}{Image representation} & \multicolumn{7}{c|}{Text Representation} &  & \multicolumn{1}{c|}{} &  & \multicolumn{1}{l|}{} \\ 
\cline{3-22}
\multicolumn{1}{l}{} &  & \rotatebox{90}{Social Network} & \rotatebox{90}{Body features} & \multicolumn{1}{c|}{\rotatebox{90}{Color}} & \multicolumn{1}{c|}{\rotatebox{90}{Texture}} & \multicolumn{1}{c|}{\rotatebox{90}{CNN}} & \multicolumn{1}{c|}{\rotatebox{90}{R-CNN}} & \multicolumn{1}{c|}{\rotatebox{90}{Siamese-CNN}} & \multicolumn{1}{c|}{\rotatebox{90}{RNN}} 
& \multicolumn{1}{c|}{\rotatebox{90}{Others}} & \multicolumn{1}{c|}{\rotatebox{90}{Pure}} 
& \multicolumn{1}{c|}{\rotatebox{90}{CNN}} & \multicolumn{1}{c|}{\rotatebox{90}{RNN}} & \multicolumn{1}{c|}{\rotatebox{90}{Word2Vec}} & \multicolumn{1}{c|}{\rotatebox{90}{Transformer}} & \multicolumn{1}{c|}{\rotatebox{90}{Graph-Embedding}} & \multicolumn{1}{c|}{\rotatebox{90}{Others}} & \multicolumn{1}{c|}{\rotatebox{90}{Celebrity/Social Infl.}} & \multicolumn{1}{c|}{\rotatebox{90}{Fit Feedback}} & \multicolumn{1}{c|}{\rotatebox{90}{Pair Feedback}} & \multicolumn{1}{c|}{\rotatebox{90}{Others}} \\ 
\hline
	
\updated{\citet{jing2023category}} 	 & 2023 	 & \multicolumn{1}{l|}{} 	 & \multicolumn{1}{l|}{} 	 &  	 &  	 & \multicolumn{1}{c|}{} 	 &  	 &  	 &  	 &  	 &  	 &  	 &  \multicolumn{1}{c|}{\cmark}  	 &  	 &  	 & \multicolumn{1}{c|}{} 	 &  	 &  	 &  	 &  &\\ 	
\hline																					
\updated{\citet{ye2023show}} 	 & 2023 	 &  	 &  	 &  	 &  	 & \multicolumn{1}{c|}{} 	 & \multicolumn{1}{c|}{} 	 &  	 &  	 &  \multicolumn{1}{c|}{\cmark}  	 &  	 &  	 &  	 & \multicolumn{1}{c|}{} 	 &  	 &  	 &  	 &   	 &  	 &  &\\ 	
\hline																					
\updated{\citet{sarkar2023outfittransformer}} 	 & 2023 	 &  	 &  	 &  	 &  	 & \multicolumn{1}{c|}{} 	 & \multicolumn{1}{c|}{} 	 &  	 &  	 &  \multicolumn{1}{c|}{\cmark}  	 &  	 &  	 &  	 & \multicolumn{1}{c|}{\cmark} 	 &  	 &  	 &  	 &   	 &  	 & & \\ 	
\hline																					
\updated{\citet{de2023disentangling}} 	 & 2023 	 & \multicolumn{1}{l|}{} 	 &  	 &  \cmark  	 &  	 &  	 &  	 &  	 &  	 &  \cmark  	 &  	 &  	 &  	 &  	 &  	 &  	 &  	 &  	 & 	 &  &\\ \hline

\updated{\citet{DBLP:conf/wacv/JandialBCCSK22}} 	 & 2022 	 & \multicolumn{1}{l|}{} 	 & \multicolumn{1}{l|}{} 	 &  	 &  	 & \multicolumn{1}{c|}{\cmark} 	 &  	 &  	 &  	 &  	 &  	 &  	 &  \multicolumn{1}{c|}{\cmark}  	 &  	 &  	 & \multicolumn{1}{c|}{} 	 &  	 &  	 &  	 &  &\\ 	
\hline																					
\updated{\citet{DBLP:conf/iclr/DelmasRCL22}} 	 & 2022 	 & \multicolumn{1}{l|}{} 	 & \multicolumn{1}{l|}{} 	 &  	 &  	 & \multicolumn{1}{c|}{\cmark} 	 &  	 &  	 &  	 &  	 &  	 &  	 &  \multicolumn{1}{c|}{\cmark}  	 &  	 &  	 & \multicolumn{1}{c|}{} 	 &  	 &  	 &  	 &  &\\ 	
\hline																					
\updated{\citet{DBLP:journals/corr/abs-2204-02473}} 	 & 2022 	 & \multicolumn{1}{l|}{} 	 & \multicolumn{1}{l|}{} 	 &  	 &  	 &  	 &  	 &  	 &  	 &  	 &  	 &  	 &  	 &  	 &  	 & \multicolumn{1}{c|}{} 	 & \multicolumn{1}{c|}{} 	 &  	 &  	 &  	 &  \\ \hline
\updated{\citet{DBLP:conf/recsys/0001MO22}} 	 & 2022 	 & \multicolumn{1}{l|}{} 	 & \multicolumn{1}{l|}{} 	 &  	 &  	 & \multicolumn{1}{c|}{\cmark} 	 &  	 &  	 &  	 &  	 &  	 &  	 &  	 &  	 &  	 &  	 &  	 &  	 &  	 &  	 &   \multicolumn{1}{c|}{\cmark} \\ 
\hline																					
\updated{\citet{DBLP:conf/sigir/SaMMLG22}} 	 & 2022 	 & \multicolumn{1}{l|}{} 	 & \multicolumn{1}{l|}{} 	 &  	 &  	 &  	 &  	 &  	 &  	 &  	 &  	 &  	 &  	 &  	 &  	 & \multicolumn{1}{c|}{} 	 & \multicolumn{1}{c|}{} 	 &  	 &  	 &  	 &  \\ \hline

\citet{DBLP:journals/corr/abs-2008-11638} 	 & 2021 	 & \multicolumn{1}{l|}{} 	 & \multicolumn{1}{l|}{} 	 &  	 & \multicolumn{1}{c|}{} 	 & \cmark 	 & \multicolumn{1}{c|}{} 	 &  	 &  	 &\cmark  	 &  	 &  	 &  	 &  	 &  	 &  	 &  	 &  	 &  	 &  	 &  \\ 
\hline																					
\citet{DBLP:journals/tmm/HidayatiGCHSWHT21} 	 & 2021 	 & \multicolumn{1}{l|}{} 	 & \cmark 	 & \multicolumn{1}{c|}{\cmark} 	 &  	 & \multicolumn{1}{c|}{\cmark} 	 &  	 &  	 &  	 &  	 &  	 &  	 &  	 &  	 &  	 & \multicolumn{1}{c|}{} 	 & \multicolumn{1}{c|}{} 	 & \multicolumn{1}{c|}{} 	 &  	 &  	 &  \\ 
\hline																					
\citet{DBLP:journals/vldb/YuHPCXLQ21} 	 & 2021 	 & \multicolumn{1}{l|}{} 	 &  	 & \cmark 	 &  \cmark	 & \multicolumn{1}{c|}{\cmark} 	 &  	 &  	 &  	 &  	 & \multicolumn{1}{c|}{} 	 &  	 &  	 &  	 &  	 &  	 &  	 &  	 & \multicolumn{1}{c|}{} 	 & \multicolumn{1}{c|}{\cmark} 	 &  \\ 
\hline																					
\citet{9354945} 	 & 2021 	 & \multicolumn{1}{l|}{} 	 &  	 &  	 &  	 & \cmark 	 &  	 &  	 & \multicolumn{1}{c|}{} 	 &  	 &  	 &  	 &  	 & \cmark 	 &  	 & \multicolumn{1}{c|}{\cmark} 	 &  	 &  	 &  	 &  	 &\cmark  \\ 
\hline																					
\updated{\citet{DBLP:conf/sigir/WenSYZN21}} 	 & 2021 	 & \multicolumn{1}{l|}{} 	 & \multicolumn{1}{l|}{} 	 &  	 &  	 & \multicolumn{1}{c|}{\cmark} 	 &  	 &  	 &  	 &  	 &  	 &  	 &  \multicolumn{1}{c|}{\cmark}  	 &  	 &  	 & \multicolumn{1}{c|}{} 	 &  	 &  	 &  	 &  &\\ 	
\hline																					
\updated{\citet{DBLP:conf/cvpr/LeeKH21}} 	 & 2021 	 & \multicolumn{1}{l|}{} 	 & \multicolumn{1}{l|}{} 	 &  	 &  	 & \multicolumn{1}{c|}{\cmark} 	 &  	 &  	 &  	 &  	 &  	 &  	 &  \multicolumn{1}{c|}{\cmark}  	 &  	 &  	 & \multicolumn{1}{c|}{} 	 &  	 &  	 &  	 &  &\\ 	
\hline																					
\updated{\citet{DBLP:conf/kdd/NestlerKHWS21}} 	 & 2021 	 & \multicolumn{1}{l|}{} 	 & \multicolumn{1}{l|}{} 	 &  	 &  	 &  \multicolumn{1}{c|}{\cmark}  	 &  	 &  	 &  	 &  	 &  	 &  	 &  	 &  	 & \multicolumn{1}{c|}{} 	 & \multicolumn{1}{c|}{} 	 &  	 &  	 &  \multicolumn{1}{c|}{\cmark}  	 &  &\\ \hline

\citet{DBLP:conf/fsdm/XingHS20} 	 & 2020 	 & \multicolumn{1}{l|}{} 	 & \multicolumn{1}{l|}{} 	 &  	 &  	 & \multicolumn{1}{c|}{\cmark} 	 &  	 &  	 & \multicolumn{1}{c|}{\cmark} 	 &  	 &  	 &  	 &  	 &  	 &  	 &  	 &  	 &  	 &  	 &  	 & \multicolumn{1}{c|}{\cmark} \\ 
\hline																					
\citet{DBLP:conf/um/WolbitschHWH20} 	 & 2020 	 & \multicolumn{1}{l|}{} 	 & \multicolumn{1}{l|}{} 	 &  	 &  	 &  	 &  	 &  	 &  	 & \multicolumn{1}{c|}{\cmark} 	 & \multicolumn{1}{c|}{} 	 &  	 &  	 &  	 &  	 &  	 &  	 &  	 &  	 &  	 & \multicolumn{1}{c|}{\cmark} \\ 
\hline																					
\citet{DBLP:conf/kdd/LiKZBG20} 	 & 2020 	 & \multicolumn{1}{l|}{} 	 & \multicolumn{1}{l|}{} 	 &  	 &  	 & \multicolumn{1}{c|}{\cmark} 	 &  	 &  	 &  	 &  	 & \multicolumn{1}{c|}{\cmark} 	 &  	 &  	 &  	 &  	 &  	 &  	 &  	 &  	 &  	 & \multicolumn{1}{c|}{} \\ 
\hline																					
\citet{DBLP:journals/tist/BanerjeeRSG20} 	 & 2020 	 & \multicolumn{1}{l|}{} 	 &  	 &  	 &  	 & \multicolumn{1}{c|}{\cmark} 	 &  	 &  	 &  	 &  	 & \multicolumn{1}{c|}{\cmark} 	 &  	 &  	 &  	 &  	 &  	 &  	 &  	 &  	 &  	 &  \\ 
\hline																					
\citet{DBLP:journals/isci/DongZKZ20} 	 & 2020 	 & \multicolumn{1}{l|}{} 	 & \multicolumn{1}{c|}{\cmark} 	 &  	 &  	 &  	 &  	 &  	 &  	 & 	 &\cmark  	 &  	 &  	 &  	 &  	 &  	 &  	 &  	 &  	 &  	 & \multicolumn{1}{c|}{\cmark} \\ 
\hline												
\citet{DBLP:conf/wacv/TangsengO20} 	 & 2020 	 &  	 & \multicolumn{1}{l|}{} 	 & \multicolumn{1}{c|}{\cmark} 	 & \multicolumn{1}{c|}{\cmark} 	 &  	 &  	 &  	 &  	 & \multicolumn{1}{c|}{\cmark} 	 & \multicolumn{1}{c|}{} 	 &  	 &  	 &  	 &  	 &  	 &  	 &  	 & \multicolumn{1}{c|}{} 	 &  	 &  \\ 
\hline																					
\citet{article} 	 & 2020 	 & \multicolumn{1}{l|}{} 	 & \multicolumn{1}{l|}{} 	 &  	 &\cmark  	 & \multicolumn{1}{c|}{\cmark} 	 &  	 &  	 &  	 &  	 & \multicolumn{1}{c|}{} 	 &  	 & \multicolumn{1}{c|}{} 	 &  	 &  	 & \multicolumn{1}{c|}{} 	 &  	 &  	 &  	 &  	 &  \\ 
\hline	
\citet{DBLP:journals/tkde/LinRCRMR20} 	 & 2020 	 &  	 &  	 &  	 &  	 & \multicolumn{1}{c|}{\cmark} 	 &  	 &  	 & \multicolumn{1}{c|}{\cmark} 	 &\cmark  	 & \cmark 	 &  	 &  	 &  	 &  	 &  	 & \multicolumn{1}{c|}{} 	 &  	 & \multicolumn{1}{c|}{} 	 & \cmark 	 &  \\ 
\hline	
\citet{DBLP:conf/kdd/ChenHXGGSLPZZ19} 	 & 2019 	 &  	 &  	 &  	 &  	 & \multicolumn{1}{c|}{\cmark} 	 & \multicolumn{1}{c|}{} 	 & \cmark 	 &  	 &  	 &  	 & \multicolumn{1}{c|}{\cmark} 	 &  	 &  	 & \multicolumn{1}{c|}{\cmark} 	 &  	 &  	 & \multicolumn{1}{c|}{\cmark} 	 &  	 &  	 &  \\ 
\hline																					
\citet{DBLP:conf/icip/PolaniaG19} 	 & 2019 	 &  	 & \multicolumn{1}{l|}{} 	 & \cmark 	 &  	 &  	 &  	 & \multicolumn{1}{c|}{\cmark} 	 &  	 &  	 &  	 &  	 &  	 &  	 &  	 &  	 & \multicolumn{1}{c|}{} 	 &  	 & \multicolumn{1}{c|}{} 	 & \multicolumn{1}{c|}{\cmark} 	 &  \\ 
\hline																					
\citet{DBLP:conf/ijcai/HouWCLZL19} 	 & 2019 	 &  	 &  	 &  	 &  	 & \multicolumn{1}{c|}{\cmark} 	 & \multicolumn{1}{c|}{} 	 &  	 &  	 & \cmark 	 &  	 &  	 &  	 &  	 &  	 &  	 & \multicolumn{1}{c|}{} 	 &  	 & \multicolumn{1}{c|}{} 	 &  	 &  \\ 
\hline																					
\citet{DBLP:journals/corr/abs-1906-05596} 	 & 2019 	 &  	 & \multicolumn{1}{l|}{} 	 &  	 &  	 &  	 & \multicolumn{1}{c|}{\cmark} 	 &  	 &  	 & \multicolumn{1}{c|}{\cmark} 	 &  	 &  	 &  	 &  	 &  	 &  	 &  	 &  	 & \multicolumn{1}{c|}{} 	 & \multicolumn{1}{c|}{\cmark} 	 &  \\ 
\hline																\citet{DBLP:conf/www/AbdullaSB19} 	 & 2019 	 & \multicolumn{1}{l|}{} 	 & \cmark 	 &  	 &  	 &  	 &  	 &  	 &  	 &  	 &  	 &  	 & \multicolumn{1}{c|}{} 	 & \multicolumn{1}{c|}{\cmark} 	 &  	 &  	 &  	 &  	 & \multicolumn{1}{c|}{\cmark} 	 & \multicolumn{1}{c|}{} 	 &  \\ 
\hline																					
\citet{DBLP:conf/www/LinRCRMR19} 	 & 2019 	 &  	 &  	 &  	 &  	 & \multicolumn{1}{c|}{\cmark} 	 &  	 &  	 &  	 &  	 &  	 &  	 &  	 &  	 & \multicolumn{1}{c|}{\cmark} 	 &  	 & \multicolumn{1}{c|}{} 	 &  	 &  	 &  	 &  \\ 
\hline																					
\citet{DBLP:journals/mta/PanDHC19} 	 & 2019 	 & \multicolumn{1}{l|}{} 	 & \multicolumn{1}{l|}{} 	 &  	 &  	 &  	 &\cmark  	 & \cmark 	 &  	 &  	 &  	 &  	 &  	 &  	 &  	 &  	 &  	 &  	 &  	 &  	 &  \\ 
\hline										
\citet{DBLP:conf/www/YinL0019} 	 & 2019 	 &  	 & \multicolumn{1}{l|}{} 	 &  	 &  	 & \multicolumn{1}{c|}{\cmark} 	 &  	 &  	 &  	 &  	 &  	 &  	 &  	 &  	 &  	 &  	 &  	 & \multicolumn{1}{c|}{} 	 & \multicolumn{1}{c|}{} 	 & \multicolumn{1}{c|}{} 	 &  \\ 
\hline			
\citet{DBLP:conf/kdd/CardosoDV18} 	 & 2018 	 &  	 &  	 &  	 &  	 & \cmark  	 &  	 &  	 &  	 & \cmark	 & \multicolumn{1}{c|}{} 	 & \cmark 	 &  	 &  	 &  	 &  	 &  	 &  	 &  	 &  	 & \multicolumn{1}{c|}{} \\ 
\hline															\citet{DBLP:conf/mod/TuinhofPH18} 	 & 2018 	 &  	 &  	 &  	 &  	 & \multicolumn{1}{c|}{\cmark} 	 &  	 &  	 &  	 & \multicolumn{1}{c|}{} 	 &  	 &  	 &  	 &  	 &  	 &  	 & \multicolumn{1}{c|}{} 	 &  	 &  	 &  	 &  \\ 
\hline
	
\citet{DBLP:conf/recsys/GuigouresHKSBS18} 	 & 2018 	 & \multicolumn{1}{l|}{} 	 &  	 &  	 &  	 &  	 &  	 &  	 &  	 &  	 &  	 &  	 & \multicolumn{1}{c|}{} 	 &  	 &  	 & \multicolumn{1}{c|}{} 	 &  	 & \multicolumn{1}{c|}{} 	 & \cmark 	 & \multicolumn{1}{c|}{} 	 &  \\ 
\hline																															\citet{DBLP:conf/mm/HidayatiHCHFC18} 	 & 2018 	 &  	 & \cmark 	 & \multicolumn{1}{c|}{\cmark} 	 &  	 & \cmark 	 &  	 &  	 &  	 &  	 &  	 &  	 &  	 &  	 &  	 & \multicolumn{1}{c|}{\cmark} 	 &  	 & \multicolumn{1}{c|}{\cmark} 	 &  	 &  	 &  \\ 
\hline	
										
\citet{DBLP:journals/corr/abs-1806-11371} 	 & 2018 	 & \multicolumn{1}{l|}{} 	 & \multicolumn{1}{l|}{} 	 &  	 &  	 &  	 &  	 &  	 &  	 & \multicolumn{1}{c|}{} 	 &  	&  	 &  	 &  	 &  	 &  	 &  	 &  	 &  	 & \multicolumn{1}{c|}{} 	 & \multicolumn{1}{c|}{\cmark} \\ 
\hline																					
\citet{DBLP:journals/ecra/HwangboKC18} 	 & 2018 	 & \multicolumn{1}{l|}{} 	 & \multicolumn{1}{l|}{} 	 &  	 &  	 &  	 &  	 &  	 &  	 &  	 & \multicolumn{1}{c|}{\cmark} 	 &  	 &  	 &  	 &  	 &  	 &  	 &  	 &  	 &  	 & \multicolumn{1}{c|}{\cmark} \\ 
\hline																																	
\citet{DBLP:conf/mm/ZhouDZZ18}  	 & 2018 	 & \multicolumn{1}{l|}{} 	 & \multicolumn{1}{l|}{} 	 &\cmark  	 & \multicolumn{1}{c|}{\cmark} 	 & \multicolumn{1}{c|}{\cmark} 	 &\cmark  	 &  	 &  	 &  	 & \multicolumn{1}{c|}{\cmark} 	 &  	 &  	 &  	 &  	 &  	 &  	 &  	 &  	 &  	 &  \\ 
\hline
										
\citet{DBLP:journals/dase/LiuDZSH18} 	 & 2018 	 &  	 &  	 &  	 & \multicolumn{1}{c|}{} 	 & \multicolumn{1}{c|}{\cmark} 	 &  	 &  	 &  	 &  	 &  	 &  	 &  	 &  	 &  	 &  	 & \multicolumn{1}{c|}{\cmark} 	 &  	 &  	 &  	 & \multicolumn{1}{c|}{} \\ 
\hline	
\citet{DBLP:journals/mta/SunCWP18} 	 & 2018 	 & \cmark 	 &  	 &  	 & \multicolumn{1}{c|}{} 	 & \cmark 	 &  	 & \multicolumn{1}{c|}{} 	 &  	 &  	 &  	 &  	 &  	 &  	 &  	 &  	 & \multicolumn{1}{c|}{} 	 & \multicolumn{1}{c|}{} 	 &  	 &  	 &  \\ 
\hline											\citet{DBLP:conf/recsys/Jaradat17} 	 & 2017 	 &  	 &  	 & \multicolumn{1}{c|}{\cmark} 	 &  	 &  	 & \multicolumn{1}{c|}{\cmark} 	 &  	 &  	 &  	 &  	 & \multicolumn{1}{c|}{} 	 &  	 &  	 &  	 &  	 & \cmark 	 & \multicolumn{1}{c|}{} 	 &  	 &  	 &  \\ 
\hline											
 \citet{DBLP:journals/sp/ZhangLSGXZTF17} 	 & 2017 	 &  	 & \cmark 	 &  	 &  	 &  	 &  	 &  	 &  	 & \multicolumn{1}{c|}{} 	 &  	 &  	 &  	 &  	 &  	 &  	 &  	 &  	 &  	 &  	 & \multicolumn{1}{c|}{\cmark} \\ 
\hline																					
\citet{DBLP:journals/corr/QianGSVS17} 	 & 2017 	 &  	 &  	 & \multicolumn{1}{c|}{\cmark} 	 & \multicolumn{1}{c|}{\cmark} 	 &  	 & \multicolumn{1}{c|}{\cmark} 	 &  	 &  	 &  	 &  	 &  	 &  	 &  	 &  	 &  	 &  	 &  	 &  	 &  	 & \multicolumn{1}{c|}{} \\ 
\hline																					
\citet{DBLP:conf/aaai/ZhaoHBW17} 	 & 2017 	 &  	 & \multicolumn{1}{l|}{} 	 &  	 &  	 &  	 &  	 & \cmark 	 &  	 &  	 &  	 & \multicolumn{1}{c|}{\cmark} 	 &  	 &  	 &  	 &  	 &  	 &  	 &  	 & \cmark 	 & \multicolumn{1}{c|}{\cmark} \\ 
\hline	        
\citet{DBLP:conf/recsys/HeinzBV17} 	 & 2017 	 &  	 &  	 &  	 & \multicolumn{1}{c|}{} 	 & \multicolumn{1}{c|}{\cmark} 	 &  	 &  	 & \multicolumn{1}{c|}{\cmark} 	 &  	 &  	 & \multicolumn{1}{c|}{} 	 &  	 &  	 &  	 & \multicolumn{1}{c|}{} 	 & \multicolumn{1}{c|}{\cmark} 	 &  	 &  	 &  	 & \multicolumn{1}{c|}{} \\ 
\hline																					
\citet{DBLP:conf/icdm/KangFWM17} 	 & 2017 	 &  	 & \multicolumn{1}{l|}{} 	 &  	 &  	 & \multicolumn{1}{c|}{} 	 &  	 & \cmark 	 &  	 &  	 &  	 & \multicolumn{1}{c|}{} 	 &  	 &  	 &  	 &  	 & \multicolumn{1}{c|}{} 	 &  	 & \multicolumn{1}{c|}{} 	 & \multicolumn{1}{c|}{} 	 &  \\ 
\hline																					
\citet{DBLP:/corr/BracherHV16} 	 & 2016 	 &  	 &  	 &  	 & \multicolumn{1}{c|}{} 	 & \multicolumn{1}{c|}{\cmark} 	 &  	 &  	 &  	 &  	 &  	 &  	 &  	 &  	 &  	 &  	 & \multicolumn{1}{c|}{\cmark} 	 &  	 &  	 &  	 &  \\ 
\hline												
\citet{DBLP:conf/sigir/McAuleyTSH15} 	 & 2015 	 &  	 & \multicolumn{1}{l|}{} 	 & \multicolumn{1}{c|}{} 	 & \multicolumn{1}{c|}{} 	 & \multicolumn{1}{c|}{\cmark} 	 &  	 &  	 &  	 & \cmark 	 &  	 &  	 & \multicolumn{1}{c|}{} 	 &  	 &  	 &  	 &  	 &  	 &  	 &\cmark  	 & \multicolumn{1}{c|}{} \\ 
\hline			
\citet{DBLP:conf/mm/HuYD15} 	 & 2015 	 &  	 & \multicolumn{1}{l|}{} 	 & \multicolumn{1}{c|}{\cmark} 	 & \cmark 	 & \multicolumn{1}{c|}{} 	 &  	 &  	 &  	 & \multicolumn{1}{c|}{} 	 & \multicolumn{1}{c|}{} 	 &  	 & \multicolumn{1}{c|}{} 	 & \multicolumn{1}{c|}{} 	 &  	 &  	 &\cmark  	 &  	 &  	 &  	 &  \\ 
\hline	
\citet{DBLP:journals/corr/BhardwajJDPC14} 	 & 2014 	 &  	 & \multicolumn{1}{l|}{} 	 & \multicolumn{1}{c|}{\cmark} 	 & \multicolumn{1}{c|}{} 	 &  	 &  	 &  	 &  	 & \multicolumn{1}{c|}{} 	 &  	 &  	 &  	 &  	 &  	 &  	 &  	 &  	 &  	 &  	 & \multicolumn{1}{c|}{} \\ 
\hline																						
								
\citet{DBLP:journals/corr/JagadeeshPBDS14} 	 & 2014 	 &  	 &  	 & \multicolumn{1}{c|}{\cmark} 	 & \multicolumn{1}{c|}{\cmark} 	 &  	 &  	 &  	 &  	 & \multicolumn{1}{c|}{} 	 &  	 &  	 &  	 &  	 &  	 &  	 &  	 &  	 &  	 &  	 &  \\ 
\hline	
\end{tabular}
}
\end{table}


\subsection{Categorization based on input}\label{subsec:fars_data_input}
By input, we refer to the data fashion RS use to train their models. They could be based on a \textit{U-I matrix} together with \textit{side information beyond the U-I matrix} (related to users and item), and contextual information. This  information determines the type of RS, according to CF, hybrid (CF+CBF) systems, and context-aware (CA) systems. In Table~\ref{tbl_feats}, we provide a classification of data/computational features used for the design of fashion RS by authors over the years.



\subsubsection{Fashion RS relying on user-item interaction data}
Recommendation models that are trained purely from U-I interaction data are known as collaborative filtering (CF). Neighborhood models are among traditional CF classes that predict the unknown U-I preferences based on the neighborhoods 
\updated{defined over} 
user-user or item-item similarities \updated{(user-based CF and item-based CF, respectively), derived from the preferences or interactions of used with products.}
Machine-learned recommendation models are powered by model-based CF, a parameterized model whose parameters are learned in the context of an optimization framework, such as Bayesian Personalized Ranking (BPR)~\cite{rendle2009bpr}. \updated{In \citet{DBLP:journals/corr/abs-1806-11371} BPR is used to address the typically vast size of the catalog in e-commerce platforms with a two-stage recommendation formulation. The first stage generates non-personalized similar candidate products, and the second stage uses a model-based CF trained with BPR to provide user-level personalization (top 50 styles).}

Matrix factorization (MF), or BPR-MF, is one of the most promising approaches to building CF. MF essentially encodes the complex relations between users and items into a lower-dimensional (latent) representation of users and items, whose dot product explains the unknown predictions~\cite{DBLP:journals/computer/KorenBV09}. \updated{\citet{DBLP:journals/ecra/HwangboKC18} propose a fashion RS that simultaneously proposes substitutes and complements based on previously purchased items, with the substitutes driving the majority of purchases according to evaluation data. The authors further consider a preference decay factor to account for the life span of a product.}

Thanks to deep neural models' ability to handle nonlinear data through nonlinear activation functions, recently \textit{deep neural CF models} have been adopted by the research community to address the linearity issue of MF-based approaches and thereby uncover a more complex relationship between users and items.
Given the challenges and nuances in the fashion domain,
most of the works incorporate additional or side information to the problem and, hence, will be discussed in the following subsections.

\subsubsection{Fashion RS relying on user-item interaction and side-information.} \label{sec:UI_side_info}
The range of information sources that RSs adopt can stretch beyond the U-I matrix, typically provided by attributes or information related to items such as the category of the user such as user gender, age, and demographics~\cite{shi2014collaborative}.  
To surface valuable recommendations, it is essential to learn from available user interactions, understand, and uncover the underlying decision factors. We categorize this side information according to items and users.


\noindent \textit{Fashion RS relying on U-I data and side-information of items.}\label{subsec_sideinfo} 
Understanding clothing provides a good platform for making recommendations~\cite{Humberto20}. State-of-the-art fashion RS employ a variety of types of content features, such as visual and textual features, as side-information to U-I interaction data. In the following, we review the most prominent features and attributes used for item representation in fashion recommendation literature~\cite{Humberto20,deldjoo2020recommender}:

\begin{figure}[!t]
    \centering
    \includegraphics[width = 0.7\linewidth]{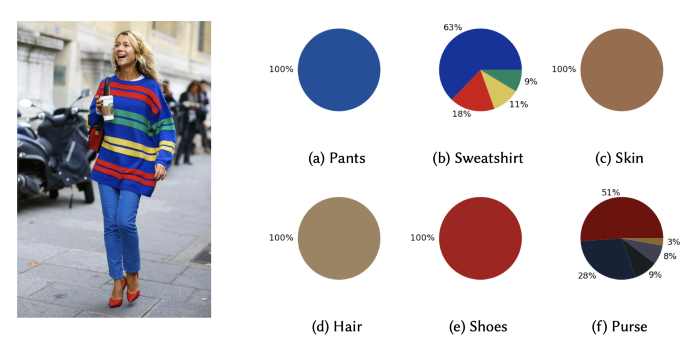}
    \caption{Color extraction example from probabilistic color modelling of clothing items. Courtesy of~\cite{al2021probabilistic}.}
    \label{fig:fashion_RS_color_extraction}
\end{figure}

\begin{enumerate}
    \item \textbf{Color.} The most common means to identify how one looks is achieved via colors, materials, and silhouettes on the body. Color and color consistency are among the most recognizable fashion clothing features that consumers often consider to decide what they want to wear. Color can be represented in several forms. For instance, as shown in Figure~\ref{fig:fashion_RS_color_extraction},~\citet{al2021probabilistic} describe a technique to segment and extract the color of a person's picture through probabilistic color modeling for clothing items. This technique can either increase or replace the data entry procedure needed to add a fashion item to electronic commerce catalogs.~\citet{DBLP:journals/corr/JagadeeshPBDS14}  
    evaluate 
    two image representations, HSV Histograms and color bag-of-word  (BoW). Their results showed that color 
is a better descriptor than textures. In~\cite{DBLP:journals/corr/QianGSVS17}, a recommendation model is proposed that uses a color map through k-means based on the value of the segmented pixel of the clothing items with an aim toward extracting a dominant color. Starting from an attribute query,~\citet{DBLP:conf/iccel/ChaeXSM18} convert this into a color vector space query; in particular, the Lab color space is used to simulate human visual perception through a palette of 269 colors.
  \item  \textbf{Brand.} 
  Product brands are a critical feature users consider when deciding among items.
  \citet{Wakita2016TowardFR} 
  define a fashion-brand recommendation system, which through a deep (feed-forward) neural network 
  has the primary purpose of predicting the user's favorite brand starting from user information, such as gender and height. \citet{DBLP:conf/recsys/Jaradat17}  propose a clothing recommendation system that 
  uses
  brands as the main feature, where the model suggests to users clothes of similar brands that 
  they have
  already purchased and that are on-trend. 
  \item \textbf{Deep Visual Features.} Fashion recommendations can be improved using image-level features extracted through a deep network, such as a CNN~\cite{DBLP:conf/fsdm/XingHS20,DBLP:conf/mm/ZhouDZZ18}. For instance, \citet{DBLP:conf/mm/ZhouDZZ18} propose a CNN implementation to address the issue of two-piece apparel matching that is suitable with current fashion trends. They merged perception and reasoning models and constructed two parallel CNNs to enable the system to recognize garment features, one for upper-body clothing and another for lower-body apparel. A hierarchical topic model incorporates the resulting information into style topics with better semantic understanding to interpret the collocation patterns. The authors present a novel learning model based on Siamese Convolutional Neural Networks (SCNN) for learning a feature transformation from clothing photos to a latent feature space expressing fashion style consistency.

   \item \textbf{Texture.} The texture describes the body and surface of a garment. It 
   has
   a direct impact on users, and is often used in recommendation systems. For example,~\citet{DBLP:conf/wacv/TangsengO20} propose an outfit recommendation system based on attributes that are ``human-interpretable,'' including texture and highlight how useful it is to use them.~\citet{DBLP:journals/corr/ZouZWLCW16} use eight different Local binary patterns (LBPs) to extract texture features from 
   pixels of an
   input image, after which a local vector is calculated for each region and encoded to obtain the final vector. \citet{DBLP:journals/corr/QianGSVS17} provide a system that uses a CNN to locate textures in an image dataset to create matches and provide recommendations. On the other hand,~\citet{DBLP:journals/corr/JagadeeshPBDS14} affirm that the color of the clothing is more important than the type of texture because the former is a stronger descriptor than the latter.
   \item \textbf{Style.} Style is associated with how 
   people intend to express themselves
   through visual elements such as accessories, clothing, hair, and other 
   aesthetic features.
   Measuring style computationally and offering personalized style-based fashion recommendations is 
   important
   for retail companies and the central focus of many research works.
   A wide range of methods has been proposed to model style in fashion recommendation, ranging from modeling the body characteristics of people with the features of clothes presented in~\citet{DBLP:journals/tmm/HidayatiGCHSWHT21}, modeling fashion substitution and complementarity with other products by learning a parametric transform of distances proposed by~\citet{DBLP:conf/sigir/McAuleyTSH15} and multi-modal representations to build trend-sensitive models within the fashion field as proposed by~\citet{DBLP:conf/mm/ZhouDZZ18}. 
   \item \textbf{Knowledge Graph (KG).}  A KG is a heterogeneous graph where nodes serve as entities and edges serve as relationships between entities. The idea of making recommendations using side information from a KG has received considerable interest in the RecSys community. Such an approach can address the issues with recommender systems and provide explanations for recommended items. For instance,~\citet{DBLP:conf/sigir/LiW0CXC20} propose a model called a Hierarchical Fashion Graph Network, which uses a graph network to model the relationship between users and items for outfit recommendations through the propagation of information based on three levels, which are items, outfit level, and user level. Another possible use of KG is related to the solution of the cold start problem of many recommendation systems, for example,~\citet{DBLP:journals/access/YanCZ19} use a model in which they build a user-item knowledge graph to discover the relationships that exist between them and solve the cold start problem. \citet{9354945}  address the problem of personalized outfit recommendation by proposing an Attentive Attribute-Aware Fashion Knowledge Graph (A3-FKG) that is used to associate various outfits with both outfit- and item-level features. \updated{\citet{DBLP:journals/isci/DongZKZ20} use a knowledge base to model the relationships between human bodies, fashion themes, and design factors using fuzzy techniques. The resulting ontology-based design knowledge base is used in an interactive, personalized design recommender system.}
   
    \item \textbf{Textual features.}  Textual features in the fashion domain can have several forms e.g., reviews \cite{DBLP:conf/sigir/ChenCXZ0QZ19}, article titles \cite{DBLP:conf/aaai/ZhaoHBW17}, article descriptions \cite{DBLP:conf/www/LinRCRMR19} and tags \cite{DBLP:/corr/BracherHV16}. \citet{DBLP:conf/aaai/ZhaoHBW17} propose a \textit{sentence model} that combines pairs of article titles to determine any compatibility of their styles.
    Textual features are integrated/modeled in the system for different purposes, for instance, for fashion generation \cite{DBLP:conf/www/LinRCRMR19}. In such a scenario, given an image of a top and a query vector description of the bottom, the fashion generator needs to generate a bottom image that matches the top image and the description as much as possible, process reviews \cite{DBLP:conf/sigir/ChenCXZ0QZ19} or attribute prediction \cite{DBLP:conf/kdd/CardosoDV18}. A shown in Table~\ref{tbl_feats}, textual information has been used to represent item content in different forms, including TF-IDF \cite{DBLP:conf/mm/ZhouDZZ18}, CNN \cite{chen2019pog, DBLP:conf/aaai/ZhaoHBW17}, Word2Vec \cite{DBLP:conf/kdd/CardosoDV18,DBLP:conf/mm/HuYD15}, transformer architecture \cite{DBLP:conf/kdd/ChenHXGGSLPZZ19}, graph-embedding \cite{DBLP:conf/recsys/GuigouresHKSBS18,DBLP:conf/mm/HidayatiHCHFC18} among others.
\end{enumerate}

Some works have tried to evaluate the quality of the features mentioned above for fashion recommendation. For example,%
~\citet{DBLP:journals/corr/JagadeeshPBDS14} compare textual and visual features or~\citet{deldjoo2021study} compare different CNN types for visual modeling of fashion items. In particular, the aim of~\citet{DBLP:journals/corr/JagadeeshPBDS14} is to make efficient use of a large amount of fashion street images. To this end, 
two
data-driven models are proposed, a deterministic fashion recommender (DFR) and a stochastic fashion recommender (SFR). The first is useful for identifying recommendations that are as objective as possible, the latter is used to model the bias of users.
An important aspect highlighted is that color is more important than the type of texture, which underlines how a simple descriptor such as color can be important.
\citet{deldjoo2021study} evaluate the quality of several visual fashion recommender systems, including VBPR, DeepStyle, ACF, and VNPR (introduced in Section~\ref{subsec:algs}) empowered by pre-trained CNN types such as Alexnet, VGG-19, and ResNet 50. Evaluation is performed on accuracy and beyond-accuracy objective and qualitative assessment of the visual similarities between pairs of images.

    

\textit{Fashion RS relying on interactions and side-information of users.} 
These approaches in the most canonical form, use data from social networks and fashion magazines and websites, in which user-user and body features (e.g., face, makeup, and size) information are extracted;


The main source of side information on users stretching beyond the U-I matrix is \textit{social networks}, which has impacted a wide range of research disciplines in RS~\cite{shi2014collaborative}. Mining social trust, influential figures, celebrities, and bloggers are among the influential factors impacting users' fashion choices. Different types of social network (SN)-based relationships have been used in the RS domain to enhance the quality of recommendations, including membership, friendship, following, and so forth. 

Different types of online platforms have been used for the purpose of advertising, marking, social, and similar activities involving fashion and garments.
\begin{itemize}
    \item \textbf{Social media.} 
With the advent of social media platforms like Instagram, the way fashion is advertised, and even the way it is designed is being re-imagined. Hundreds of thousands of people share images of their "outfit of the day," which prompts a flood of comments and queries from other users. Fashion brands can now use social media platforms such as Instagram to grow their businesses. Today, a single post by a fashion influencer receives much attention allowing many followers to know more about the brands and styles of the clothes that users mostly follow to understand which brands and models users follow the most, or to catch the trends of the moment or seasonality~\cite{jaradat2018dynamic,DBLP:conf/recsys/Jaradat17}.     

    \item \textbf{Fashion magazines and beauty websites.} It is especially common for celebrities' styles to be viewed as fashion inspirations as they pay fashion stylists to assist them with their wardrobe choices, which allows them to visually alter their actual body figure. In~\cite{DBLP:conf/mm/HidayatiHCHFC18}, the authors extract a list of essential female celebrities from prominent Fashion Magazines such as Vogue, Glamour, Marie Claire. In addition, a suite of body parameters is extracted from the body measurement website associated with celebrities. The dataset was completed using female celebrity photos taken from Google search Engines. Polyvore\footnote{\url{www.polyvore.com}} was another good example of a fashion website, where users could build and post information about the fashion outfits reflecting their taste. These fashion outfits feature a wealth of multimodal data, including photographs and descriptions of fashion products, the number of times the outfit has been liked, and the outfit's hash tags. Researchers have applied this data to a variety of fashion-related tasks (cf. Section \ref{sec:eval}).

\end{itemize}

\citet{DBLP:journals/mta/SunCWP18} propose a personalized clothing recommendation system, which takes into account both users’ social circles and fashion style consistency of clothing products. Fashion style consistency refers to the fact that two clothing items, e.g., tops and skirts, can be visually different, but as long as they belong to the same style (e.g., sport, street, casual), the user may be likely buy them; hence fashion style consistency is an essential element for the design of clothing RS.
To this end, the proposed approach considers three factors in the proposed clothing RS, namely interpersonal influence, personal interest, and interpersonal interest similarity; this was motivated by previous research works such as \cite{DBLP:journals/tkde/QianFZM14}. 
In particular, five types of matrices are built by mining the social data available and other sources of information: representing user-user social influence, representing the user-user similarity of interests, representing user-clothing similarity, representing clothing-clothing fashion style similarity, and representing user-clothing ratings.  Afterward, a probabilistic matrix factorization (PMF) framework is used to integrate the above observed matrices and recommend suitable clothing products to users by casting the problem in an optimization setting. Evaluation is carried out on real-world datasets collected from Moguije\footnote{\url{http://mogu-inc.com/}}, a Chinese website for social fashion blending an SN with online shopping options.

\subsubsection{Context-aware fashion RS}\label{subsec:context}  Context-aware (CA) models improve CF by including contextual information into the model to account for the current information need of the user~\cite{deldjoo2020recommender,aggarwal2016ensemble}. Context awareness, i.e., understating the user's situational or contextual aspects, is essential in improving the user experience. Several taxonomies have been identified to classify context, e.g., based on computing environment, user environment, and physical environment or based on time, location, and social information (user's friends, social circles), see~\citet{deldjoo2021content} for more information. For the scope of this survey, we choose a pragmatic approach and organize CA Fashion RS according to the types of context frequently used in the fashion literature, namely \textit{spatio-temporal context} (e.g., GPS coordinates, location, place of interest, time, season), \textit{multimedia and physiological context} (e.g., free text, image or the user body/face features), and  \textit{affective context} (e.g., mood or emotion).
\vspace{1.5mm}

\noindent \textit{Spatio-temporal context.}
The \textit{temporal} and \textit{geographic} characteristics of fashion data influence human preferences. As for temporal aspects, first, fashion garments are sensitive to \textit{seasonal} changes. Retail markets may face a decrease in preference over time after fashion products are released.
Second, fashion \textit{style trends} change as time progresses. For instance, trends can appear cyclically, with styles re-emerging after decades of absence, 
while others emerge
and
swiftly disappear. Understanding how style choices \textit{vary with time}, and thereby what constitutes fashionable trends, is of crucial importance to the fashion retail business. 

\citet{DBLP:conf/www/HeM16} propose an approach that involves learning the time evolution with which users decide to buy items through their implicit feedback, such as purchase history, clicks, and bookmarks. These features are used to define a predictor, helpful in approximating how much users interact with articles in a particular epoch. Dynamic FDNA proposed in \cite{DBLP:conf/recsys/HeinzBV17} extends the static FDNA model with the inclusion of time-of-sale information. FDNA combines attributes and visual items using a neural architecture. The dynamic FDNA takes as input the timestamp of the purchase associated with a customer, and 
using existing information about the customer
(purchase history so far), 
determines their current style.
The dynamic model has the main advantage of providing recommendations for short-term customer intent. \citet{inbook} propose a framework, 
which provides via a weather API
recommendations for the exact day when the user needs them.\\




\noindent \textit{Multimedia and body/facial-related context.} The term \dquotes{multimedia context} refers to situations in which the system requires a user to provide a multimedia item (e.g., an image of a fashion item or an image plus textual query) in order to initiate the recommendation process (e.g., recommending a garment that complement an item the user is wearing). 

\begin{itemize}
    \item \textit{Context = image.}  Images are an important visual tool for users to communicate with a fashion recommender system. They could be used as input to the system to retrieve complementary or matching fashion products. For instance, given an image of a fashion item (e.g., \dquotes{jeans}), one can identify matching fashion products (e.g., \dquotes{tops}) complement an item the user is considering~\citep{DBLP:journals/corr/JagadeeshPBDS14,DBLP:conf/mm/ZhouDZZ18}. An application in the context of fashion is \dquotes{upselling in e-commerce} when an online shopper with an item in her shopping basket is prompted to buy more things that match the one they already have in their shopping carts.
    

    \item \textit{Context = image + text.} \updated{In the fashion domain, multi-modal dialogue systems that can accurately recommend or generate matching fashion items based on a user's query image and textual description can potentially revolutionize how people shop for clothes. These systems can help users easily find the perfect outfit by providing personalized and accurate recommendations based on their preferences~\cite{DBLP:conf/sigir/WenSYZN21,DBLP:conf/cvpr/LeeKH21,DBLP:conf/www/LinRCRMR19,DBLP:conf/wacv/JandialBCCSK22,DBLP:conf/iclr/DelmasRCL22}. The problem addressed in these works can be typically formally started as given a query image and a textual description of modifications; the objective is to train a model that can learn an Image-Text representation that closely aligns with the visual representation of the target ground-truth image. \citet{DBLP:conf/sigir/WenSYZN21} introduce a Comprehensive Linguistic-Visual Composition Network (CLVC-Net) that utilizes fine-grained local-wise and global-wise composition modules, along with a mutual enhancement module to retrieve the target image. The system can handle simple text modifications such as \dquotes{change it to long sleeves} and abstract visual property adjustments like \dquotes{change the style to professional}. \citet{DBLP:conf/cvpr/LeeKH21} propose a Content-Style Modulation (CoSMo) algorithm that introduces two modules - the content and style modulators - to achieve desired modifications to the reference image feature for image retrieval with text feedback. \citet{DBLP:conf/iclr/DelmasRCL22} propose an efficient model that consists of two modules ( cross-modal and visual retrieval) with text-guided attention layers, each handling one modality of the query. These modules are trained jointly. These systems hold great potential for developing \textbf{multi-modal dialogue systems}, and other tasks such as \textbf{virtual try-on} to effectively understand and respond to human queries involving text and images.}

    \item \textit{Context =  body/facial-related attributes.} \updated{The user's physical appearance, including their face or body, can provide valuable contextual information to a recommendation system, which can then suggest cosmetic products \cite{DBLP:journals/sp/ZhangLSGXZTF17,DBLP:journals/mta/Chung14}, clothing \cite{DBLP:journals/tmm/HidayatiGCHSWHT21,DBLP:conf/mm/HidayatiHCHFC18}, and hair-style recommendations \cite{Liu2013WowYA,Yang2012HS}. For instance, \citet{DBLP:journals/tmm/HidayatiGCHSWHT21} propose a framework that uses both body characteristics and clothing style to provide size recommendations. Meanwhile, in~\cite{DBLP:conf/mm/HidayatiHCHFC18}, a system requires users to provide information about their body, such as their shape, weight, height, bust, waist, and hip measurements, which is then correlated with photos of celebrities to suggest a style that matches the user's preferences.}

\end{itemize}

In Table \ref{tab:fars_input}, we provide reference points to research works that explicitly use specific multimedia contexts (such as text and images) for fashion item information access (recommendation or search). 

\begin{table}[h]
\centering
\caption{Classification of fashion RS based on the input and output of the system. 
    Compl.~pair/outfit, refers to settings where \textbf{complementary} relationships between fashion items 
    are considered,
    either for pairwise 
    matches
    or outfit recommendation}
\label{tab:fars_input}
\resizebox{\linewidth}{!}{%
\begin{tabular}{c|c|c|c} 
\hline
\multicolumn{1}{c|}{\textbf{Class}} & \textbf{Input} & \textbf{Output} & \multicolumn{1}{c}{\textbf{Research work}} \\ 
\hline
\multirow{2}{*}{\begin{tabular}[c]{@{}c@{}}\textbf{Classical }\\\textbf{ fashion RS}\end{tabular}} & – & item & \cite{DBLP:conf/recsys/HeinzBV17,DBLP:conf/recsys/GuigouresHKSBS18,DBLP:conf/ijcai/HouWCLZL19,DBLP:conf/recsys/Jaradat17,DBLP:journals/ecra/HwangboKC18,DBLP:journals/mta/SunCWP18,DBLP:journals/sp/ZhangLSGXZTF17} \\ 
\cline{2-4}
 & – & pair./outfit & \cite{DBLP:conf/wacv/TangsengO20,DBLP:conf/kdd/ChenHXGGSLPZZ19,DBLP:conf/www/AbdullaSB19,DBLP:conf/mm/HuYD15,DBLP:journals/ecra/HwangboKC18,DBLP:journals/mta/SunCWP18} \\ 
\hline
\multicolumn{1}{l|}{} & image & item & \cite{DBLP:conf/kdd/CardosoDV18,DBLP:conf/mod/TuinhofPH18,DBLP:journals/corr/abs-1806-11371,DBLP:conf/icdm/KangFWM17,DBLP:journals/corr/BhardwajJDPC14,DBLP:conf/www/YinL0019,DBLP:journals/jvcir/ZhouMZZSQC19,DBLP:journals/corr/QianGSVS17,DBLP:journals/corr/JagadeeshPBDS14} \\ 
\cline{2-4}
\multicolumn{1}{l|}{} & image & pair/outfit & \cite{DBLP:conf/icip/PolaniaG19,DBLP:journals/corr/abs-1906-05596,DBLP:journals/tkde/LinRCRMR20,DBLP:journals/corr/BhardwajJDPC14,DBLP:conf/www/YinL0019,DBLP:journals/jvcir/ZhouMZZSQC19,DBLP:journals/dase/LiuDZSH18,DBLP:journals/mta/SunCWP18,DBLP:conf/aaai/ZhaoHBW17,DBLP:conf/sigir/McAuleyTSH15} \\ 
\cline{2-4}
\multirow{3}{*}{\begin{tabular}[c]{@{}c@{}}\textbf{Contextual }\\\textbf{ fashion RS}\end{tabular}} & image+text & item & \cite{DBLP:conf/mm/ZhouDZZ18,DBLP:conf/sigir/WenSYZN21,DBLP:conf/cvpr/LeeKH21} \\ 
\cline{2-4}
 & image+text & compl. pair/outfit & \cite{DBLP:conf/www/LinRCRMR19,DBLP:conf/mm/ZhouDZZ18} \\ 
\cline{2-4}
 & body/face & item & \cite{DBLP:conf/mm/HidayatiHCHFC18,DBLP:journals/mta/Chung14} \\ 
\cline{2-4}
\multicolumn{1}{l|}{} & body/face & compl. pair/outfit & \cite{DBLP:journals/sp/ZhangLSGXZTF17} \\ 
\hline
\multirow{2}{*}{\begin{tabular}[c]{@{}c@{}}\textbf{GAN-based }\\\textbf{ fashion image gen.}\end{tabular}} & – & item & \cite{DBLP:conf/icdm/KangFWM17, DBLP:conf/aaai/ShihCLS18,article2, Yang2018FromRT} \\ 
\cline{2-4}
 & – & compl. pair/outfit & \cite{DBLP:journals/corr/abs-1906-05596, DBLP:conf/eccv/HuynhCTA18} \\
\hline
\end{tabular}
}
\end{table}

\section{Algorithms for Fashion RS}\label{subsec:algs}
For the purpose of this survey, we focus on the two most prominent approaches for fashion RS, (i) visually-aware model-based CF (cf. Section~\ref{subsub:VRS}), and (ii) generative fashion recommendation models (cf. Section~\ref{subsec:GAN_FA_rec}). Finally, we briefly discuss other fashion RS approaches in (cf. Section~\ref{subsec:others}).

\subsection{Visually-aware model-based CF}
\label{subsub:VRS}

The fashion industry has a wide catalog of diverse items and a high rate of change or return as a result of market dynamics and customer preferences. This results in a lack of purchase data, which complicates the use of standard recommender systems. In addition, it is difficult to compute similarities across products when exact and extensive product information is not provided. Computer vision research is increasingly being used to address the issues outlined above. Given the visual and aesthetic nature of fashion products, a growing body of research addresses tasks like localizing fashion items, determining their category and attributes, or establishing the degree of similarity to other products. For instance, \citet{DBLP:journals/corr/QianGSVS17} use a CNN to segment and locate elements from the complicated backgrounds of street-style images. \citet{DBLP:conf/ijcai/HouWCLZL19} use a semantic segmentation approach to derive region-specific attribute representations. 




\setlength{\parindent}{15pt} \textbf{VBPR (2016)~\cite{DBLP:conf/aaai/HeM16}}
The visual Bayesian Personalized Ranking (VBPR) method is built upon BPR and extends it by incorporating a latent content-based preference factor. The core predictor in VBPR is given by
\begin{equation}
  \label{eq:VBPR}
   \hat{x}_{u, i} =  p_u^Tq_i + \theta^T_u \theta_i
\end{equation}
where $\theta_i  \in \mathbb{R}^{K}$ is typically designed to represent the visual signal of an item, and  $\theta_u  \in \mathbb{R}^{K}$ represents the user preference on the visual dimension. \citet{DBLP:conf/aaai/HeM16} built the VBPR predictor according to
\begin{equation}
  \label{eq:VBPR_var}
   \hat{x}_{u, i} =  p_u^Tq_i + \theta^T_u \ E \ \underbrace{\Phi_f(\mathbf{Img}_i)}_\text{$f_i$}
\end{equation}
in which $f_i = \Phi_f(\mathbf{Img}_i) \in \mathbb{R}^{F \times 1}$ represents the latent feature of item $i$ (typically extracted from a pretrained CNN) and $\Phi_f$ denotes the feature extractor. $E \in \mathbb{R}^{K \times D}$  linearly transforms the high-dimensional feature $F$ into a lower-dimensional (e.g., $D =$ 20) named the visual space. In~\cite{DBLP:conf/aaai/HeM16}, the authors show that VBPR improves the performance of BPR-MF and popularity-based recommenders for datasets chosen from Amazon fashion and Amazon phones. 


\setlength{\parindent}{15pt} \textbf{DeepStyle (2017)~\cite{DBLP:conf/sigir/LiuWW17}.} While VBPR has demonstrated effectiveness, it tends to learn the visual characteristics of items based on their category rather than their specific style (such as casual, aesthetic, or formal). Creating a style-CNN requires labeled data, which can be a time-consuming task that requires expert knowledge in the field of fashion. However, DeepStyle presents an innovative solution to this problem by using a U-I matrix to learn the style of items. The approach involves modeling the item as a combination of its style and category, where removing the category information is achieved by modeling the style as the difference between the item and the category (\texttt{style} = \texttt{item} - \texttt{category}). This transformation effectively converts the VBPR predictor into a dedicated style-CNN.
 \begin{equation}
        \hat{x}_{u, i} = p_u^Tq_i + p_u^T(\mathbf{E}f_i - l_i)
    \end{equation}
where $l_i$ is a latent factor that embodies the categorical information of item $i$.

\setlength{\parindent}{15pt} \textbf{ACF (2017)~\cite{DBLP:conf/sigir/ChenZ0NLC17}} \updated{Attentive collaborative filtering (CF) integrates an attention network to consider the varying levels of attention that users give to different components within multimedia content, such as the regions in an image or frames in a video when making recommendations. This approach is more effective than traditional visual recommendation systems that do not account for such nuances in user behavior}
\begin{equation}
\label{eq:ACF}
\hat{x}_{u, i} = \bigg(p_u + \sum_{k \in \mathcal{I}_u^+} a(u, k)v_k\bigg)^Tq_i
\end{equation}
in which $a(u, k)$ denotes user $u$'s preference degree toward item $k$, and $v_k$ is the attentive latent factor for item $k$, and $q_i$ is the basic item vector in the MF model.

\setlength{\parindent}{15pt} \textbf{DVBPR (2017)~\cite{DBLP:conf/icdm/KangFWM17}} \updated{VBPR utilizes pre-trained CNNs, such as ResNet50, to extract visual information. However, these networks are trained for image classification tasks rather than recommendation tasks, and they come from different domains. In contrast, DVBPR is an end-to-end model that trains the visual feature extractor and the preference prediction module simultaneously in a pairwise manner}
     \begin{equation}
       \label{eq:DVBPR}
        \hat{x}_{u, i} = \theta_u^T\Phi_e(\mathbf{Img}_i)
    \end{equation} 
where the product $i$ is associated with an image denoted as $\mathbf{Img}_i$, and its item content embedding is represented by $\Phi_e(\mathbf{Img}_i)$. The CNN model used to generate this embedding is trained directly in a pairwise manner.

Other relevant techniques in this area are TVBPR by ~\citet{DBLP:conf/www/HeM16}, which incorporates seasonality and temporal changes into the VBPR model, and CO-BPR by~\citet{DBLP:conf/www/YinL0019}, which recommends compatible outfits based on BPR. These methods draw inspiration from traditional content-based recommendation techniques, with the primary challenge being the higher dimensionality and computational costs associated with visual data. To address this, the image representations are projected into lower-dimensional spaces (as with VBPR) or handled through end-to-end (pixel-level) representations.

A general classification of CV in Fashion RS is presented by~\citet{cheng2021fashion}, according to (i) fashion detection (landmark detection, fashion parsing),  (ii) fashion analysis (attribute prediction, style learning, popularity prediction), and (iii) fashion synthesis (style transfer, pose estimation).


\subsection{Generative Fashion Recommendation Models}

\label{subsec:GAN_FA_rec}

Most conventional RS are not suitable for application in the fashion domain due to unique characteristics hidden in this domain. 
Recently, GAN-based models have been used for fashion generation and fashion recommendation with promising performance, for example \citet{DBLP:journals/corr/abs-1906-05596} use an enhanced conditional GAN to generate bottom items that can fit with the top items received in input by their framework. GANs gain their power exploiting their \textit{generative power}, allowing them to synthesize realistic-looking fashion items \cite{DBLP:journals/corr/abs-2005-10322}. This aspect can inspire customers' and designers' aesthetic appeal/curiosity and motivate them to explore the space of potential fashion styles.


\setlength{\parindent}{15pt} [\textbf{CRAFT}] \citet{DBLP:conf/eccv/HuynhCTA18} address the problem of recommending complementary fashion items based on visual features by using an adversarial process that resembles a GAN and uses a conditional feature transformer as $\mathcal{G}$ and a discriminator $\mathcal{D}$. One main distinction between this work and the prior literature is that the $\langle$input, output$\rangle$ pair for $\mathcal{G}$ are both features (here features are extracted using pre-trained CNNs~\cite{DBLP:conf/aaai/SzegedyIVA17}), instead of $\langle$image, image$\rangle$ or hybrid types such as $\langle$image, features$\rangle$ explored in numerous previous works~\cite{DBLP:conf/iccv/ZhuFULL17,DBLP:conf/cvpr/VolpiMSM18}. This would allow the network to learn the relationship between items directly on the feature space, spanned by the features extracted. The proposed system is named complementary recommendation using adversarial feature transform (CRAFT) since in the model, $\mathcal{G}$ acts like a feature transformer that---for a given query product image $q$---maps the source feature $s_{q}$ into a complementary target feature $\hat{t}_{q}$ by playing a min-max game with $\mathcal{D}$ with the aim to classify fake/real features. For training, the system relies on learning the co-occurrence of item pairs in real images. In summary, the proposed method does not generate new images; instead, it learns how to generate features of the complementary items conditioned on the query item. 

\setlength{\parindent}{15pt} [\textbf{DVBPR}] Deep visual Bayesian personalized ranking (DVBPR)~\cite{DBLP:conf/icdm/KangFWM17} is 
one of the first works 
to
exploit the \textit{visual generative power of GANs} in the fashion recommendation domain. It aims to generate clothing images based on user preferences. Given a user and a fashion item category (e.g.,~tops, t-shirts, and shoes), the proposed system generates new images---i.e.,~clothing items---that are consistent with the user's preferences. The contributions of this work are two-fold: first, it builds an end-to-end learning framework based on the Siamese-CNN framework. Instead of using the features extracted in advance, it constructs an end-to-end system that turns out to improve the visual representation of images. Second, it uses a GAN-based framework to generate images that are consistent with the user's taste. Iteratively, $\mathcal{G}$ learns to generate a product image integrating a \textit{user preference maximization objective}, while $\mathcal{D}$ tries to distinguish generated images from real ones. Generated images are quantitatively compared with real images using the preference score (mean objective value), inception score~\cite{DBLP:conf/nips/SalimansGZCRCC16}, and opposite SSIM~\cite{DBLP:conf/icml/OdenaOS17}. This comparison shows an improvement in preference prediction in comparison with non-GAN based images. At the same time, the qualitative comparison demonstrates that the generated images are realistic and plausible, yet they are quite different from any images in the original dataset---they have standard shape and color profiles, but quite different styles.

\setlength{\parindent}{15pt} [\textbf{MrCGAN}] \citet{DBLP:conf/aaai/ShihCLS18} propose a compatibility learning framework that allows the user to visually explore candidate \textit{compatible prototypes} (e.g., a white t-shirt and a pair of blue-jeans). The system uses metric-regularized conditional GAN (MrCGAN) to pursue the item generation task. It takes as 
input a projected prototype (i.e.,~the transformation of a query image in the latent ``compatibility space''). It produces as the output a synthesized image of a compatible item (the authors consider a compatibility notion based on the complementary of the query item across different catalog categories). Similar to the evaluation protocol in~\cite{DBLP:conf/eccv/HuynhCTA18}, the authors conduct online user surveys to evaluate whether their model could produce images that are perceived as compatible. The results show that MrCGAN can generate compatible and realistic images under compatibility learning settings compared to baselines.

\setlength{\parindent}{15pt} [\textbf{Yang et al.} \& \textbf{$c^+$GAN}] \citet{Yang2018FromRT} address the same problem settings of MrCGAN~\cite{DBLP:conf/aaai/ShihCLS18} by proposing a fashion clothing framework composed of two parts: a clothing recommendation model based on BPR combined with visual features and a clothing complementary item generation based GAN. Notably, the generation component takes as input a piece of clothing recommended in the recommendation model and generates clothing images of other categories (i.e.,~tops, bottom, or shoes) to build a set of complementary items. The authors follow a similar qualitative and quantitative evaluation procedure as DVBPR~\cite{DBLP:conf/icdm/KangFWM17} and further propose a \textit{compatibility index} to measure the compatibility of the generated set of complementary items. A similar approach has also been proposed in $c^+$GAN~\cite{DBLP:journals/corr/abs-1906-05596}, to generate a bottom fashion item paired with a given top.

\subsection{Other Fashion Recommender System Algorithms}\label{subsec:others}
Depending on the task, fashion item and outfit recommendation, and size recommendation as we identified in Section~\ref{subsec:fars_task}, a variety of other models have been used in the studied research work. For example, RNNs~\cite{DBLP:journals/tkde/LinRCRMR20,DBLP:conf/kdd/CardosoDV18}, graph-modeling~\cite{DBLP:conf/sigir/McAuleyTSH15,DBLP:conf/sigir/LiW0CXC20,DBLP:conf/kdd/ChenHXGGSLPZZ19}, two-input (Siamese) CNNs~\cite{DBLP:conf/icdm/KangFWM17,DBLP:conf/icip/PolaniaG19,DBLP:conf/www/YinL0019,DBLP:conf/mm/ZhouDZZ18,DBLP:conf/aaai/ZhaoHBW17}, using attention mechanisms~\cite{DBLP:journals/tkde/LinRCRMR20} and so forth. Detailed discussion on these approaches 
is left as a
future direction. 

\textbf{~3. Feature Encoding.}
\updated{The output of the last step is a series of feature vectors $\{x_i\}$. 
This step now represents each feature $x_i$ using a codebook, typically by mapping $x_i$ into one or several feature coding vector(s)
$v_i \, , \,  (v_i \in \mathrm{R}^K)$, where $K$ is the number of codewords determined based on training data. The codewords may be predefined or learned (e.g., by clustering the training data with $k$-means). 
The size and characteristics of the codebook impact the subsequent item description and thereby its discrimination power~\cite{liu2019bow}. In the case of images, common feature encoding methods include vector quantization (VQ), $k$-means, sparse coding, vector of locally aggregated descriptors (VLAD), and Fisher vectors (FV).}

\textbf{~4. Pooling.} \updated{The goal of this step is to produce a fixed-length global feature representation 
that statistically summarizes the properties of the item's segments 
by aggregating the feature coding vectors $\{v_i\}$. The most common pooling methods include average pooling and max pooling.} 

\subsection{Data mining and post processing}

\updated{Fashion data mining heavily relies on the quality of item representations, in particular the segmentation and the extracted features. The task of data mining entails utilizing the extracted features (item representations) in a middle step to classify semantic metadata (e.g., image category) and similar annotation task. Also, integration of textual data with visual item representations is discussed in this step. We refer to this type of \textit{multimodal fusion}. We discuss these two steps in the following:}

\textbf{~1. Image annotation/classification.} \

\textbf{~2. multimodal fusion.} \

\textbf{Classification based on the multi-modal fusion approach.}
\begin{itemize}
    \item \textbf{FDNA} [8] combines attributes and visual items using a neural architecture. The system passes EGC to an DNN and produces a single vector. It also feeds a CNN images of fashion items and produces a single vector, too. Fusion is done by mixing them into a vector of dimension 2d, which is called FDNA.
    \item \textbf{Dynamic FDNA} [24] is an evolution of FDNA model, it is based on LSTM cells. While the implementation of the FDNA model used articles as input, the dynamic FDNA is customer based. It takes in input the timestamp of the purchase \textit{i} associated with customer \textit{k}, and thanks to the information about the customer \textit{k} who saved in own memory (purchase history so far), it sends out the current customer style of \textit{k}.
    \item \textbf{FARM} [69] combines visual and textual information as input queries using a transformer. It is used a CNN as a vector encoder to extract information from them, which are used to generate a completely image that match with the input query. To improve the recommendation, there are extracted information from a candidate image and then it is evaluated with the input query.
   \updated{ \item  In \cite{DBLP:conf/mm/HidayatiHCHFC18}, the authors provide a \textbf{Fashion Recommendation system on the clothing style for personal nody shape}, they use a dataset consisting of images of famous female characters in addition to a set of parameters regarding their body(type of body shape, dress size, height, weight,...). Their framework is based on graphs built from the visual elements of the images and information about their body shape.}
   \updated{\item Attribute prediction and classification are very important tasks, especially when you have a lot of clothing images available but are not well labeled. In this sense, the authors \cite{DBLP:conf/kdd/CardosoDV18} propose to merge visual and textual content through a \textbf{multi-modal multi-task architecture} to predict attributes while \textbf{DSFCH} \cite{DBLP:journals/dase/LiuDZSH18} provides to classify high-level attributes of clothing items like an hash codes. }
\end{itemize}

\updated{ \textit{Categorization based on data sources involved.} :}
\begin{itemize}
    \item Image + text
    \item Image + text + interaction data
\end{itemize}

\updated{ \textit{Categorization based on the method.} :}

\textbf{~3. Feature/region importance measurement.} \updated{One of the important operations that can be considered in FARS's design is to measure the importance of a feature at hand or clothing regions (parts). The goal of this feature/region assessment can be to explain the judgments and to improve the recommendation quality. For instance,~\citet{DBLP:conf/wacv/TangsengO20} measure the positive and negative influence of some interpretable visual features for outfit recommendation and use it for explaining recommendations.}

\begin{figure}
    \centering
    \includegraphics[width = 0.54\linewidth]{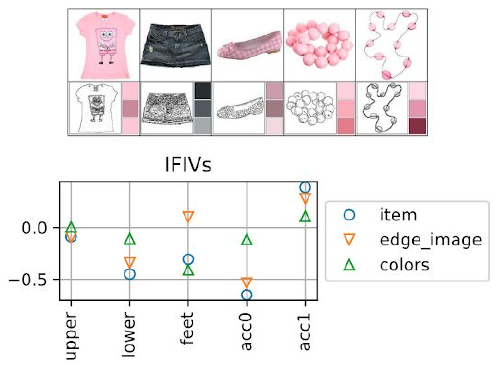}
    \caption{An example of visual explanation offered to users for fashion recommended items. Courtesy of~\cite{DBLP:conf/wacv/TangsengO20}}
    \label{fig:explain}
\end{figure}

\section{Evaluation and Datasets}
\label{sec:eval}
Evaluation of fashion recommendations is a very complex task. It has to be adapted to the specific recommendation task (see Section~\ref{subsec:fars_task}) and consider the inputs of the algorithms (Section~\ref{subsec:fars_data_input}).
Hence, in the next section we present several goals that aim to be achieved when deciding for a specific evaluation setting (Section~\ref{subsec:eval_goal}), whereas Section~\ref{subsec:eval_pers} presents the most important perspectives analyzed in the literature; finally, Section~\ref{subsec:avail_data} shows and discusses currently available datasets.

\subsection{Evaluation goal}\label{subsec:eval_goal}
The goal of the evaluation is usually linked with the task the recommendation is intended to satisfy.
However, we may find proposals where the algorithm is evaluated under different goals. For example, an outfit recommender can be evaluated only for that goal by ranking lists of outfits, but a more simple setting could be devised by using the recommender to score different outfits, in order to validate whether the algorithm is capable of discriminating those outfits that occur in the groundtruth.

In the following, we present the most important goals we have identified in the literature:

\begin{itemize}
     \item \textbf{Goal 1 -- Outfit generation}: The goal of this task is, given a fashion item $x_q$ (e.g., skirt) representing the user's current interest, find the best item $x\in\mathcal{I}$ (e.g., shirt) or the fashion outfit $F \in \mathcal{I}$ (e.g., shirt, pants, hat) that goes/go well with the input query. The input $x_q$ can be specified as a textual query (e.g., what outfit goes well to this mini skirt?) or a visual query (photo of a skirt). 
    
    When $x_q$ is a visual query, the task is recognized as a \textbf{visual retrieval} task in the community of information retrieval (IR) or \textbf{contextual recommendation} in the community of RS ($x_q$ is a multimedia context representing a user's interest -- see e.g. \cite{DBLP:journals/csr/KaminskasR12} for a definition of multimedia context (cf. Section \ref{subsec:context}).
   

    \item \textbf{Goal 2 -- Outfit Recommendation}: 
    This goal is related to the fashion and outfit recommendation task, where a set of objects is recommended to the user at once, 
    by maximizing a utility function that measures the suitability of recommending a fashion outfit to a specific user.
    
    \item \textbf{Goal 3 -- Pair recommendation}: This goal is a simplification of the outfit recommendation goal when $N=2$. This is typically performed as \textit{top-bottom} or \textit{bottom-up} recommendation. In this task, given clothes related to the upper part of the body, the aim is to predict the possible lower part and vice versa. For instance, in a mobile setting, the user may take a photograph of a single piece of clothing of interest (e.g., tops) related to the upper part of the body and look for the bottom (e.g., trousers or skirts) from a large collection that best match the tops~\cite{DBLP:journals/corr/JagadeeshPBDS14}. This typically requires a collection of pairs of top and bottom images for compatibility modeling. With advancements in computer vision, in some works, automated top-bottom region detection from photograph images has been proposed~\cite{iwata2011fashion}. 
    
    \item \textbf{Goal 4 -- Fill In the Blank} (FITB): This goal 
    is used in a setting where we are
    given an incomplete outfit $F^-$ (e.g., shirt, pants, accessories) with a missing item (e.g., shoes), and the method must find the best missing fashion item $x \in \{x_1, x_2, x_3 ,x_4\}$ from multiple choices where $F = \{F^-,x\}$ has fashion items, which are compatible visually~\cite{han2017learning}. 
    This is a convenient scenario in real life, e.g., a user wants to choose a pair of shoes to suit his pants and jacket. 
    
    \item \textbf{Goal 5 -- Outfit compatibility prediction}: This goal is focused on, given a complete outfit $F \in \mathcal{F}$, predicting a compatibility score that best describes the composition of an outfit. All items in an outfit are compatible when all fashion items have similar style and go well together \cite{tangseng2020toward,han2017learning}.
    This task is helpful since users may create their own outfits and wish to decide if they are compatible and trendy.
\end{itemize}

\subsection{Evaluation perspective}\label{subsec:eval_pers}
Besides the goal that is considered when evaluating a recommender system, we may find authors that take different perspectives when considering what is a good recommendation.
Although most of the literature in the fashion domain has been focused on building more accurate prediction and recommendation models, as in other recommendation settings, recent research has gone beyond this perspective and analyzed the explanations that should be presented to the user, the images that go with the recommendations, and even social aspects that may reinforce biases in the system.
We describe these perspectives below.

\begin{itemize}
    \item \textbf{Evaluate the recommendation}:
        Most papers perform offline evaluation, where classification or ranking metrics are very popular (precision, nDCG, AUC); in this sense, business metrics like CTR or purchase percentage are less common.
        See Table~\ref{tab:eval} for a summary.
        It should be noted that, depending on the recommendation task, other concepts, such as compatibility, might be measured to assess how well the user recognizes the recommendation.
        \updated{Recent approaches like \cite{DBLP:journals/corr/abs-2204-02473} propose comparative recommendations focused on product discovery tasks. Other works explicitly address the context of the recommendation, in particular, the \textit{scene} the outfit is expected to be considered \cite{ye2023show}, natural-language feedback \cite{DBLP:conf/recsys/0001MO22}, or maximum shopping budget \cite{DBLP:journals/tist/BanerjeeRSG20}. }
    
    \item \textbf{Evaluate an explanation}: 
    Understanding recommendation explanations is generally a difficult task, since theoretically it can be evaluated only by real users.
    These approaches can be however costly and produce subjective judgements, however recent techniques aim to produce interpretable recommendations, instead of providing an explicit explanation in natural language.
    Examples of research works where this perspective is considered for the fashion domain are \cite{yang2019interpretable,wu2020visual}. 
    
    \item \textbf{Evaluate generated images}:
    When generating images, 
    researchers 
    typically measure three complementary metrics \cite{DBLP:conf/icdm/KangFWM17}: preference (how much each user would be interested in the recommended items), image quality using the inception score as a proxy, and diversity (where the visual similarity between every pair of returned images is computed).
    Sometimes, qualitative evaluation is also included. For example, \citet{DBLP:conf/www/LinRCRMR19} show several examples of generated items and discuss their validity. A similar qualitative evaluation is observed in \cite{yang2018recommendation}.
    
     \item \textbf{Evaluate social perspectives}: \citet{DBLP:conf/hcse/BrandG20} highlights a very important aspect of today's society, gender equality, in detail the attention is placed on the price differences that emerge between men's and women's products. 
     Other works like \cite{10.4018/JITR.2020070107} argue whether users perceive in the same way recommendations generated by humans or by services exploiting Artificial Intelligence, in terms of trustworthiness and acceptance of the recommendations.
     In a similar line, the diversity of the recommendations (where the product exposure is higher) has been linked to a higher purchase rate and amount, although it may depend on the type of customer \cite{10.1145/3340531.3412687}.
     In any case, special care must be taken by automatic fashion recommender systems to not reinforce social biases and perpetuate long-observed objectifications \cite{TIGGEMANN201345,10.1111/j.1460-2466.2012.01667.x}, hence, more research on these lines should be derived from the ML and RecSys communities.
\end{itemize}
    
\noindent From the summary presented in Table \ref{tab:eval}, we conclude that the most prominent perspective remains the first one (\textit{evaluate the recommendation}). Nonetheless, some recent papers pay attention to other perspectives, such as image generation and social perspectives. This can be observed in the table by checking how many papers report image quality/diversity metrics, or metrics beyond accuracy, such as diversity or some kind of explanation measurements. The general trend, hence, is that offline evaluation with classical metrics (error, classification, and ranking) are prominent, but the community is working towards introducing others, more focused on alternative dimensions, such as business metrics or those related to social and explanation perspectives.

\begin{table}[htbp]
\centering
\caption{The list of most commonly used datasets in the Fashion RS literature.}
\label{tab:datasets}
\resizebox{\linewidth}{!}{%
\begin{tabular}{c|c|c|c|c} 
\hline
\multicolumn{1}{c|}{\textbf{Dataset}} & \textbf{Reference} & \textbf{Availability} & \textbf{Size} & \multicolumn{1}{c}{\textbf{Area}} \\ 
\hline
~Amazon Reviews & \href{http://jmcauley.ucsd.edu/data/amazon/links.html}{Link} & Yes & 140 million interactions & Generic \\ 
\hline
~ DeepFashion~ ~ ~ & \cite{liu2016deepfashion} & No & 8 hundred thousand images & Generic \\ 
\hline
~ ~Exact Street2Shop & \href{http://www.tamaraberg.com/street2shop/}{Link} & Yes & - & Street and shop photos \\ 
\hline
~ ~ ~ExpFashion~ ~ ~ & \cite{DBLP:journals/tkde/LinRCRMR20} & No & 200k images and 1 million comments & Item \\ 
\hline
~ ~Fashion IQ~ ~ ~ & \cite{DBLP:conf/cvpr/WuGGARGF21} & Yes & 77.6K images and textual descriptions & Dresses, shirts, tops\&tees \\ 
\hline
~ ~FashionVC~ ~ ~ & \cite{10.1145/3123266.3123314} & No & 20~thousand images & Outfits by fashion experts \\ 
\hline
~ ~ Fashion-136K~ ~ ~ & \cite{DBLP:journals/corr/JagadeeshPBDS14} & No & - & Fashion models \\ 
\hline
IQON3000 & \cite{DBLP:conf/mm/SongHLCXN19} & Yes & 300K outfits, by 3.5K users & Outfits \\ 
\hline
Polyvore & \cite{DBLP:conf/mm/HuYD15} & No & - & Item photos \\ 
\hline
Pog & \href{https://github.com/wenyuer/POG}{Link} & Yes & 4 millions images and descriptions & Item and Outfit \\ 
\hline
~ ~ StyleReference~ ~ ~ & \cite{DBLP:journals/tmm/HidayatiGCHSWHT21} & No & 2~thousand pictures & Body features \\ 
\hline
~ ~ ~ ~Style4BodyShape~ ~ ~ & \cite{DBLP:conf/mm/HidayatiHCHFC18} & No & 3150 celebrity records & Female celebrities \\
\hline
\end{tabular}
}
\end{table}

\subsection{Available datasets}\label{subsec:avail_data}

Table~\ref{tab:datasets} summarizes a list of the most frequently used datasets, together with information on their availability, size, and the area to which the data pertain. They are discussed in more detail next.


%

 \begin{itemize}
  \item \textbf{Amazon Reviews dataset.} 
  Amazon is the world's largest marketplace for
  selling several categories of products,
  including fashion products. The collection consists of about 140 million records, including product reviews, ratings, and product information.
  
   \item \textbf{DeepFashion.} DeepFashion contains over 800K photos annotated with 50 categories, 1K attributes, and clothing landmarks (each image contains four to eight landmarks), as well as over 300K image pairs. They are classified according to several contexts, such as the store, street photo, and consumer. 

    \item \textbf{Exact Street2Shop.} The images in this collection fall into two categories:
    \textit{(a)} street images, which are photographs of individuals wearing various garments and taken in spontaneous, unplanned moments;
    \textit{(b)} shop photos, photographs of clothing products from online clothing businesses, known as \dquotes{shop photos}.
    
     \item \textbf{ExpFashion.} This dataset was gathered from a collection of Polyvore images. They created new outfits given an item from current outfits and placed them in the dataset, starting with 1,000 seed outfits and expanding to a total of 100,000 outfits.
    

    \item \updated{\textbf{Fashion IQ.} This dataset is commonly used as a benchmark for image retrieval with natural language feedback. It contains 77.6K images of fashion products over 3 categories: Dress, Toptee, and Shirt, with 46.6K images in the training and around 15.5K images for both the validation and test sets.}
     
     \item \textbf{FashionVC.} Dataset consisting of the outfits of Polyvore's most fashion experts. Using a seed set of Polyvore  popular outfits, the researchers identified 256 fashion experts, and then retrieved historical outfit from them.
   
    \item \textbf{Fashion-136K.} For this dataset, photographs of fashion models were gathered from the web. All photographs will have both top and bottom garments in the same image.

    \item \updated{\textbf{IQON3000.} This dataset is composed of garment images and metadata. Garments are grouped by outfit and are associated to different users. These data were collected from IQON, a Japanese fashion community website, in which members could mix fashion items in order to create new outfits.}
   
      \item \textbf{Polyvore.} A social commerce platform that allows users to exchange items and utilize them in image collages called Sets.
      As a dataset for many fashion recommendation systems (e.g., \citet{DBLP:conf/mm/HuYD15} have defined 150 sets of users), Sets are employed because they include only the products and a clean background, which substantially facilitates extraction of object properties.
   
    \item \textbf{POG.} The dataset contains data retrieved from the Taobao website; specifically, the clicks on the most popular items and outfits were extracted, and each one is paired with a record including the image's background, title, and category.
    
   
    \item \textbf{StyleReference.} There are 2,160 photos of apparel in this collection. The photographs were sourced from a number of major fashion publications style magazines (e.g., Elle and Vogue).

     \item \textbf{Style4BodyShape.}  In this dataset, it is possible to find three different kinds of information: \textit{(1)} a list of the most stylish 3,150 female celebrities, who are well-known for their refined sense of fashion; \textit{(2)} female celebrity body measurements, including dress, bust, waist, and hip measures, derived from a website that compiles this data, and \textit{(3)} images of 270 fashionable celebrities crawled via Google search engine.

\end{itemize}

In summary, we notice that most of the reported datasets are not publicly available. This is a big hurdle to promote comparable and reproducible research in this domain.
Moreover, we observe the area to which the data belongs to is heterogeneous, as not two datasets share the same area except the three datasets classified as \textit{Generic}.
Finally, even though most of these datasets are quite large (reaching millions of interactions), some of them are very small, including few thousands of images, which might be insufficient information for some methods to learn interesting patterns from them.

\begin{table}[h]
\caption{Common datasets and evaluation metrics used in the fashion recommender literature.
}\label{tab:eval}
\scalebox{0.70}{
\begin{tabular}{l l c c c c c c c c c c c c c c l c}
\toprule
\small{Authors} & \small{Year} & \begin{tabular}[c]{@{}c@{}}\small{} \\ \small{}\end{tabular} & \multicolumn{12}{c}{\small{Evaluation}} &\small{} &\begin{tabular}[c]{@{}c@{}}\small{Clothing} \\ \small{type}\end{tabular} & \small{Datasets} \\ \cline{4-14}
& & &\small{type} &\multicolumn{11}{c}{\small{metric}} & & & \\ \cline{5-14}
& & & &\multicolumn{5}{c}{\small{offline metrics}} & &\multicolumn{5}{c}{\small{online/qual./business}} & & & \\ \cline{5-9} \cline{11-15}
& & &  &\rotatebox{90}{\small{Error}} &\rotatebox{90}{\small{Clf.}} &\rotatebox{90}{\small{Rank}} &\rotatebox{90}{\small{Beyond}} &\rotatebox{90}{\small{Expl.}} & &\rotatebox{90}{\small{Survey}} &\rotatebox{90}{\small{business}} &\rotatebox{90}{\small{image quality}} &\rotatebox{90}{\small{image diversity}} &\rotatebox{90}{\small{.}} & & & \\ \toprule
\small{Tangseng et al.~\cite{DBLP:conf/wacv/TangsengO20}} & \small{2020} &\small{} &\small{offline} & &\small{\cmark} & & &\cmark & & & & & & &  &\small{generic} &\small{PO} \\ \hline
\small{Lin et al.~\cite{DBLP:journals/tkde/LinRCRMR20}} & \small{2020} &\small{} &\small{offline} & &\small{\cmark} &\small{\cmark} & & \small{\cmark}& & & & & & & &\small{top-bottom} &\small{FashionVC} \\ \hline
\small{Chen et al.~\cite{DBLP:conf/kdd/ChenHXGGSLPZZ19}} & \small{2019} &\small{} &\small{off+online} & &\small{\cmark} & & & & &  &\small{\cmark} & & & & &\small{generic} &\small{POG} \\ \hline
\small{Polania et al.~\cite{DBLP:conf/icip/PolaniaG19}} & \small{2019} &\small{} &\small{offline} & & &\small{\cmark} & & & & &\small{\cmark} & & & & &\small{generic} &PO \\ \hline
\small{Hou et al.~\cite{DBLP:conf/ijcai/HouWCLZL19}} & \small{2019} &\small{} &\small{offline} & &\small{\cmark} &\small{\cmark} & & & & & & & & & &\small{generic+shoe} &\small{Amazon} \\ \hline
\small{Kumar et al.~\cite{DBLP:journals/corr/abs-1906-05596}} & \small{2019} &\small{} &\small{offline} & &\small{\cmark} & &\small{\cmark} & & & & & \small{\cmark}&\small{\cmark} & & &\small{top-bottom} &\small{Bing} \\ \hline
\small{Zhou et al.~\cite{DBLP:journals/jvcir/ZhouMZZSQC19}} & \small{2019} &\small{} &\small{online} & &\small{} & & & & & & & & & & &\small{generic} &\small{NA} \\ \hline
\small{Pan et al.~\cite{DBLP:journals/mta/PanDHC19}} & \small{2019} &\small{} &\small{offline} & &\small{\cmark} & & & & &\small{\cmark} & & & & & &\small{generic} &\small{NA} \\ \hline
\small{Lin et al.~\cite{DBLP:conf/www/LinRCRMR19}} & \small{2019} &\small{} &\small{offline} & &\small{\cmark} &\small{\cmark} & & & &\small{\cmark} & & & & & &\small{top-bottom} &\small{ExpFashion} \\ \hline
\small{Yin et al.~\cite{DBLP:conf/www/YinL0019}} & \small{2019} &\small{} &\small{offline} & &\small{\cmark} & & & & & &&\small{\cmark}  & & & &\small{generic} &Amazon \\ \hline
\small{Liu et al.~\cite{DBLP:journals/dase/LiuDZSH18}} & \small{2019} &\small{} &\small{offline} & &\small{\cmark} & & & & & & & & & & &\small{top-bottom} &FarFetch \\ \hline
\small{Abdulla et al.~\cite{DBLP:conf/www/AbdullaSB19}} & \small{2019} &\small{} &\small{off+online} & &\small{} & & & & & & & & & & &\small{generic} &\small{NA} \\ \hline
\small{Hwangbo et al.~\cite{DBLP:journals/ecra/HwangboKC18}} & \small{2018} &\small{} &\small{online} & & & & & & & &\small{\cmark} & & & & &\small{generic} &NA \\ \hline
\small{Cardoso et al.~\cite{DBLP:conf/kdd/CardosoDV18}} & \small{2018} &\small{} &\small{offline} & &\small{\cmark} & & & & & & & & & & &\small{generic} &ASOS \\ \hline
\small{He et al.~\cite{DBLP:journals/corr/abs-1810-02443}} & \small{2018} &\small{} &\small{online} & &\small{} & & & & & & & &\small{\cmark} & & &\small{generic} &\small{PO} \\ \hline

\small{Sun et al.~\cite{DBLP:journals/mta/SunCWP18}} & \small{2018} &\small{} &\small{offline} & \small{\cmark}&\small{\cmark} & & & & & & & & & & &\small{top-bottom} &MouJIE \\ \hline
\small{Cardoso et al.~\cite{DBLP:conf/kdd/CardosoDV18}} & \small{2018} &\small{} & \small{offline}& &\small{\cmark} & & & & & & & & & & &\small{generic} &\small{ASOS} \\ \hline
\small{Tuinhof et al.~\cite{DBLP:conf/mod/TuinhofPH18}} & \small{2018} &\small{} &\small{offline} & &\small{\cmark} & & & & & & & & & & &\small{top-bottom} &\small{CrowdFlower} \\ \hline
\small{Hidayati et al.~\cite{DBLP:conf/mm/HidayatiHCHFC18}} & \small{2018} &\small{} &\small{off+online} & &\small{} &\small{\cmark} & & & &\small{\cmark} & & & & & &\small{Woman Dress} &\small{Style4Body} \\ \hline
\small{Zhou et al.~\cite{DBLP:conf/mm/ZhouDZZ18}} & \small{2018} &\small{} &\small{} & &\small{} &\small{\cmark} & & & & & & & & & &\small{top-bottom} &DeepFashion \\ \hline
\small{Guigoures et al.~\cite{DBLP:conf/recsys/GuigouresHKSBS18}} & \small{2018} &\small{} &\small{offline} & &\small{\cmark} & &\small{\cmark} & & & & & & & & &\small{shoes} &NA \\ \hline
\small{Agarwal et al.~\cite{DBLP:journals/corr/abs-1806-11371}} & \small{2018} &\small{} &\small{offline} & &\small{\cmark} &\small{\cmark} & & & & & & & & & &\small{Men T-Shirt} &\small{Myntra} \\ \hline
\small{Heinz et al.~\cite{DBLP:conf/recsys/HeinzBV17}} & \small{2017} &\small{} &\small{offline} & &\small{\cmark} & & & & & & & & & & &\small{generic} &\small{Zalando} \\ \hline
\small{Kang et al.~\cite{DBLP:conf/icdm/KangFWM17}} & \small{2017} &\small{} &\small{offline} & &\small{\cmark} & & & & &\cmark & &\cmark &\cmark & & &\small{shoe+top-bot.} &\small{Amazon} \\ \hline
\small{Zhang et al.~\cite{DBLP:journals/sp/ZhangLSGXZTF17}} & \small{2017} &\small{} &\small{offline} & &\small{\cmark} & & & & & & & & & & &\small{-} &NA \\ \hline
\small{Jaradat et al.~\cite{DBLP:conf/recsys/Jaradat17}} & \small{2017} &\small{} &\small{offline} & &\small{} & & & & & & & & & & &\small{-} &\small{Zalando+Instagram} \\ \hline

\small{Qian et al.~\cite{DBLP:journals/corr/QianGSVS17}} & \small{2017} &\small{} &\small{offline} & &\small{\cmark} &\small{\cmark} & & & & & & & & & &\small{top-bottom} &\small{-} \\ \hline
\small{Zhao et al.~\cite{DBLP:conf/aaai/ZhaoHBW17}} & \small{2017} &\small{} &\small{offline} & &\small{\cmark} & & & & & & & & & & &\small{generic} &Amazon+Tao \\ \hline
\small{Bracher et al..~\cite{DBLP:/corr/BracherHV16}} & \small{2016} &\small{} &\small{online} & &\small{} & & & & & & & & & & &\small{generic} &\small{Zalando} \\ \hline
\small{McAuley et al.~\cite{DBLP:conf/sigir/McAuleyTSH15}} & \small{2015} &\small{} &\small{offline} & &\small{\cmark} & & & & & & & & & & &\small{-} &Amazon \\ \hline
\small{Hu et al.~\cite{DBLP:conf/mm/HuYD15}} & \small{2015} &\small{} &\small{offline} & &\small{} &\small{\cmark} & & & & & & & & & &\small{generic} &\small{PO} \\ \hline

\small{Wang et al.~\cite{DBLP:journals/thms/WangZKC15}} & \small{2015} &\small{} &\small{online} & & & & & & & \small{\cmark}& & & & & &\small{-} &\small{-} \\ \hline
\small{Bhardwaj  et al.~\cite{DBLP:journals/corr/BhardwajJDPC14}} & \small{2014} &\small{} &\small{online} & &\small{} & & & & &\small{\cmark} & & & & & &\small{top-bottom} &\small{Fashion-136K} \\ \hline
\small{Jagadeesh et al.~\cite{DBLP:journals/corr/JagadeeshPBDS14}} & \small{2014} &\small{} &\small{online} & &\small{} & & & & &\small{\cmark} & & & & & &\small{top-bottom} &\small{Fashion-136K} \\ \hline


\midrule
\multicolumn{18}{l}{\textbf{Challenges} IE: Interpretability/Explanation}  \\ \multicolumn{18}{l}{\textbf{Metrics} Err.: Error-based accuracy (RMSE, MAE), Clf.: Classification metrics (Precision, Recall, F1, Accuracy, AUC)} \\
\multicolumn{18}{l}{RA: Rank-aware accuracy (MRR, MAP, NDCG), Beyond: Beyond accuracy metric (Novelty, Coverage, Diversity)}  \\ 
\multicolumn{18}{l}{Expl.: Explanation related evaluation}  \\ 
\multicolumn{18}{l}{\textbf{Datasets} PO: Polyvore, NA: Anonymous}\\ 
\bottomrule
\end{tabular}}
\end{table}
%



\section{Conclusions}

\label{sec:con}
In this survey, we have analyzed and classified the RS  that function in a specific vertical market: clothes and fashion goods. 
In particular, we have introduced a taxonomy of fashion recommender systems, which categorizes them according to the task (e.g., item, outfit, size recommendation, explainability among others), and type of side information (users, items, context).
We have also identified the most important evaluation goals (outfit generation, outfit recommendation, pair recommendation, fill in the blank, and outfit compatibility prediction) and perspectives (evaluate the recommendation, the explanation, the generated images, or the social perspectives) exploited by the community, together with the most common datasets and evaluation metrics.
This domain presents a unique collection of challenges and sub-problems pertinent to the development of successful recommender systems.

\noindent \textbf{Datasets.}  We may recall how data collections began with simple \dquotes{harvested} datasets and progressed to more \dquotes{curated} datasets in subsequent years (e.g., images from fashion models rather than e.g., co-purchase data). It is interesting to consider which of these methods is preferable and where the field is headed in terms of datasets. Even though harvested datasets are easy to collect and correspond well to real prediction tasks (e.g., purchase estimate), collected datasets are noisy; they may also not reflect the real semantics of visual preferences, compatibility, or other aspects of a user's experience. In contrast, curated datasets may not match the distribution of real data; or \dquotes{models} may not represent the preference dimensions of regular users, or the datasets may actually be contrary to one's goals. For instance, a marketing image may try to promote a pair of shoes by pairing it with a \dquotes{boring} (non-distracting) outfit; thus curated data may not be any more \dquotes{real} than harvested data (cf. Section~\ref{subsec:avail_data}).



\noindent \textbf{Tasks.} While traditional task in fashion RS research  involved purchase or co-purchase prediction tasks, recent systems focus on combinatorial outputs (e.g. outfits and wardrobes) or even generative tasks (cf. Section~\ref{subsec:fars_task}). One needs to think whether this complexity is necessary. Do complicated non-pairwise functions actually exist in fashion semantics, or are pairwise functions adequate to represent the underlying decision factors? Is the increased complexity of combinatorial models worth the investment? This may require us to examine the long-term viability of simpler models in comparison to more complicated ones (cf. Section~\ref{subsec:fars_task}).

\noindent \textbf{Interpretation.} What exactly do visually aware models \dquotes{learn}? Are they truly capturing fine-grained fashion semantics, or are they simply learning trivial factors (e.g., categories) from fashion data? (cf. Section~\ref{subsec:expl}) Visual recommender systems are mostly used to cope with noisy, sparse, cold-start datasets including visual data, but it is unclear that we have made much progress toward acquiring \dquotes{real} fashion semantics in the manner of a designer. If our objective is just to predict purchases, this may be irrelevant; if our objective is to fulfill the function of fashion designers, how can we develop more representative datasets or tasks? (cf. Section~\ref{subsec:expl})

\section{Future Outlook}
\updated{
\noindent We anticipate that the following areas will receive future attention, and be the subject of future surveys: \vspace{1mm}
}

\updated{
\noindent \textbf{Forecasting.} The majority of fashion recommendations are focused on predicting the \dquotes{present} based on previous interactions and trends, i.e., what the user will do \emph{right now} based on their history. How might such models be utilized to make predictions regarding the fashion trends of the future? This aspect can be linked to popularity forecasting in the fashion domain since trendy items will likely be popular \cite{10.1145/2668107,Bandari_Asur_Huberman_2021}.
}

\updated{
\noindent \textbf{Ethics.}  Fashion recommendation raises a number of ethical considerations. For instance, how certain factors like gender~\cite{DBLP:conf/hcse/BrandG20}, race, body shape, or religion, to name a few, result in stereotypical recommendations. A fashion recommendation system, for instance, may be biased towards recommending clothing items that are traditionally worn by men or women, thereby ignoring the fashion tastes of people who do not conform to these traditional gender roles. Besides the ethical considerations, such bias and unfairness can have other severe consequences, including decreased retail sales, dissatisfied customers, and degradation of the brand image (see also \cite{deldjoo2023fairness,amigo2023unifying}). With the goal of rare product discovery, \citet{DBLP:conf/sigir/SaMMLG22} explore how to diversify the recommendations in the context of luxury fashion, while \citet{DBLP:conf/kdd/NestlerKHWS21} use a Bayesian model to reduce the size and fit-related product returns, hence reducing the risk of personal issues such as “it’s not me” or other body image issues.
}

\updated{
The ethics of \dquotes{fast fashion} is also worth considering. Fast fashion is a term that refers to the manufacturing process, which means that the consumer may have to choose between purchasing something cheaper but manufactured under dubious conditions or paying more (or even having a very limited selection) for \textit{sustainable} and \textit{ethically} superior clothing. 
}

\updated{
\noindent \textbf{Adversarial robustness and privacy.} Protecting the security of ML/RS models and user data. In the fashion recommendation domain, adversarial attacks are a significant concern as attackers can manipulate the system's output and deceive users \cite{deldjoo2021survey}. One common form of such an attack is to manipulate the images of clothing items, making them appear different from their actual appearance. Another form of attack involves adding adversarial perturbations (noise) to the images that are invisible to the naked eye and promoting or demoting certain items or item categories in the recommendation list for purposes such as marketing presentations. Consequently, users may be directed to clothing items they may not have otherwise considered, which could result in dissatisfaction and a reduction in retail sales \cite{treu2021fashion, anelli2021study}. An example of an attack is presented by \citet{treu2021fashion}, which uses a fashion image to generate an adversarial texture applied to the clothing regions of the target image, preventing person segmentation from working correctly.
}




\updated{
\noindent \textbf{Deployability.} What are the practical considerations in terms of deploying models from academia? Perhaps complexity and being naturally black-box models (relating to the preceding discussion about interpretation) are important aspects for consideration given that the majority of the techniques described are based on neural networks. Furthermore, good performance on a small academic dataset may fail to reproduce at an Internet-size scenario, which compounded with the lack of interpretability may result in unforeseen problems when a new model is released to the public~\cite{reese2016tay, nieva2015google}. Properly testing systematically and at scale, and how to better explain and understand model predictions, is thus a relevant area of research.
}



\updated{
\noindent \textbf{Multi-modal and cross-modal representations.} A multi-modal fashion
RS can be trained to recognize the semantic aspects of clothing items, such as their color, style, and material, and combine this information with visual data, such as images of the clothing items, to improve the product representations and ultimately provide more accurate recommendations \cite{deldjoo2021multimedia}. However, it is costly to scale the training data required for all the fashion concepts with industry-level accuracy. However, recent advances in multimodal text-and-image models such as CLIP~\cite{radford2021learning} or ALBEF~\cite{li2021align} have shown that it is possible to achieve a remarkable level of semantic understanding using only weakly related internet text and image data. Future research will probably target how to use similar approaches for fashion RS.
}

\updated{
Similarly, cross-modal text-to-image generative models, such as DALL-E~\cite{ramesh2022hierarchical} and Stable Diffusion~\cite{https://doi.org/10.48550/arxiv.2112.10752}, could be incorporated into fashion AI systems to improve the accuracy of recommendations. These models can be trained to generate images
of clothing items based on textual descriptions or retrieve textual descriptions of clothing items based on visual data, which can provide more accurate and diverse recommendations for users.
}

\updated{
\noindent \textbf{Iterative and multi-modal recommendation.} As conversational systems progress and gain wide adoption, fashion RS will be able to utilize multi-turn interactions~\cite{deldjoo2021towards,yu2019visual}, as well as multiple forms of input, to refine and improve the recommendations to the customer. Furthermore, different pieces of information provided by the customer may be contradictory or modify each other, and in future fashion RS will need to exhibit robustness to more informal forms of communication. Works such as \cite{wen2021comprehensive}, \cite{delmasartemis}\updated{, or \cite{DBLP:conf/recsys/0001MO22}}, among others, start to tap into this area, but more research will be needed to make such systems practical, and train them at scale.
}

\updated{
\noindent \textbf{Virtual try-on.} Augmented Reality (AR) is becoming commonplace in mobile camera apps, and in consequence it is very likely that customers will start expecting it in e-commerce settings. Some product vertical like beauty or eyewear, where the requirements are more aligned with current commercial AR technology, are early adopters~\cite{wang2022augmented}. Clothing virtual try-on will require more research before it becomes useful, but it will create opportunities to make fashion RS more personalized~\cite{jayamini2021use}. 
}

\noindent \textbf{Causal view and interpretability.} A final challenge for fashion AI and recommendation is to take a causal approach to fashion recommendation, which entails understanding the causal relationships between data and the effect of features on informed decision-making. For instance, a fashion AI model can be trained to understand the causal relationships between a user's clothing preferences and their body type, age, or lifestyle, and then use this knowledge to provide personalized recommendations that are more likely to suit the user's needs and preferences. This problem is related to interpretability of fashion RS predictions, a topic common to many of the future research areas mentioned above. Works such as~\cite{packer2018visually, DBLP:conf/ijcai/HouWCLZL19, tangseng2020toward} make inroads on this problem, but more research will be needed to be able to explain the recommendation of a fashion RS based on the input customer data.

\noindent \textbf{Integration with generative AI and large language models (LLMs).} Finally, as generative AI and large language models (LLMs) continue to advance, their roles in recommender systems--particularly in the field of fashion--are becoming increasingly significant \cite{wang2023generative,hou2023large,fan2023recommender}. These models can operate in both uni-modal and multi-modal capacities. For example, in the fashion industry, LLMs can generate outfit recommendations by understanding the compatibility between various fashion items in a nuanced manner. In multi-modal scenarios, these models can combine both text and visual data to offer more personalized choices. For instance, a user could interactively decide what they would like to wear by inputting both textual and visual preferences into the system. Moreover, these models can also tackle common challenges in recommender systems such as sparsity and the cold-start problem through prompt-engineering techniques such as few-shot and zero-shot learning \cite{nazary2023chatgpt,deldjoo2023fairnessChat}.

\bibliographystyle{ACM-Reference-Format}
\bibliography{refs}

\end{document}